\documentclass[aps,prc,letterpaper,twoside,tightenlines,nofootinbib,showpacs
,preprint,superscriptaddress]{revtex4-1}
\usepackage{setspace}
\usepackage{epsfig,float}
\usepackage{amsmath}
\begin{document}

%\begin{spacing}{1.3}
\newcommand{\Ima}{\textrm{Im}}
\newcommand{\Rea}{\textrm{Re}}
\newcommand{\mev}{\textrm{ MeV}}
\newcommand{\gev}{\textrm{ GeV}}
\newcommand{\rb}[1]{\raisebox{2.2ex}[0pt]{#1}}

\title{Baryon states with hidden charm in the extended local hidden gauge approach}

\author{T. Uchino}
\email{uchino@ific.uv.es}
\affiliation{
Departamento de F\'{\i}sica Te\'orica and IFIC, Centro Mixto Universidad \\de Valencia-CSIC, Institutos de Investigaci\'on de Paterna, Apartado 22085, 46071 Valencia, Spain}

\author{Wei-Hong Liang}
\affiliation{Department of Physics, Guangxi Normal University, Guilin, 541004, People's Republic of China}
\author{E. Oset}
\affiliation{
Departamento de F\'{\i}sica Te\'orica and IFIC, Centro Mixto Universidad \\de Valencia-CSIC, Institutos de Investigaci\'on de Paterna, Apartado 22085, 46071 Valencia, Spain}

\date{\today}

\begin{abstract}

The $s$-wave interaction of $\bar{D}\Lambda_c,~\bar{D} \Sigma_c,~\bar{D}^*\Lambda_c,~\bar{D}^*\Sigma_c$ and $\bar{D}\Sigma_c^*,~\bar{D}^*\Sigma_c^*$, is studied within a unitary coupled channels scheme with the extended local hidden gauge approach. In addition to the Weinberg-Tomozawa term, several additional diagrams via the pion-exchange are also taken into account as box potentials. Furthermore, in order to implement the full coupled channels calculation, some of the box potentials which mix the vector-baryon and pseudoscalar-baryon sectors are extended to construct the effective transition potentials. As a result, we have observed six possible states in several angular momenta. Four of them correspond to two pairs of admixture states, two of $\bar{D}\Sigma_c-\bar{D}^*\Sigma_c$ with $J=1/2$, and two of $\bar{D}\Sigma_c^*-\bar{D}^*\Sigma_c^*$ with $J=3/2$. Moreover, we find a $\bar{D}^*\Sigma_c$ resonance which couples to the $\bar{D}\Lambda_c$ channel and one spin degenerated bound state of $\bar{D}^*\Sigma_c^*$ with $J=1/2,5/2$.

\end{abstract}

\pacs{}

\maketitle

\section{Introduction}

In this work, possible nucleon resonances with negative parity are studied, with the energy around $4200 \sim 4400$ MeV, which are found as molecular states of one anti-charmed meson, $\bar{D}$ or $\bar{D}^*$, and a charmed baryon. Here the local hidden gauge approach with an extension to SU(4) is utilized, as done in Refs.~\cite{wumolina1,wumolina2,xiaojuan,xiaooset} to deal with the hadronic interaction involving heavy quarks. Within the framework of the local hidden gauge approach, the meson-baryon scattering proceeds via the vector-exchange \cite{hidden1,hidden2,hidden4} (see also Ref.~\cite{hideko} for practical rules) and the interaction kernel projected over $s$-wave results in the form of the Weinberg-Tomozawa interaction \cite{Tolos:2005ft,Lutz:2005vx,angelsmizu,Tolos:2007vh,wumolina1,wumolina2,wuzou}. The heavy quark spin symmetry (HQSS) is known to play an important role in the heavy quark sector \cite{isgur,neubert,manohar,wise} and has been utilized to investigate the hadron spectroscopy in the high energy region \cite{GarciaRecio:2008dp,Flynn:2011gf,GarciaRecio:2012db,Romanets:2012hm,Guo:2013xga,xiaojuan, carmen,olena}. Yet, it was also found in Ref.~\cite{xiaojuan} that the extension of the local hidden gauge approach automatically fulfils the rules of HQSS.

Compared to the light quark sector, more importance should be placed on the vector-pseudoscalar admixture because of the small mass difference between them, namely $m_{D^*}-m_{D} \sim 140$ MeV in the charm sector while $m_{K^*}-m_{K} \sim 500$ MeV. In the meson-baryon scattering, the mixing of vector-baryon ($VB$) and pseudoscalar-baryon $(PB)$ sectors proceeds via the pion-exchange induced by the $\bar{D}\bar{D}^*\pi$ vertex, which is one of the ingredients of the local hidden gauge approach \cite{hidden1,hidden2,hidden4,javier,liang,liang2}. When dealing with pion exchange, the on-shell factorization of the potential, which is done for vector exchange in the chiral unitary approach, is not possible because of the momentum dependence of the pion exchange. This is why the dynamics of pion exchange is included by means of a box diagrams, mediated by pion exchange, with the excitation of the corresponding meson-baryon intermediate states \cite{javier,liang}. The pion exchange between pseudoscalar and vector mesons is accompanied by the contact Kroll-Ruderman term \cite{javier,Carrasco:1989vq} which breaks the spin degeneracy of the vector-baryon states in the absence of this interaction \cite{liang2}.

In addition to the vector-pseudoscalar mixing term, another pion exchange contribution which stems from the anomalous $\bar{D}^*\bar{D}^*\pi$ vertex is also taken into account \cite{liang2}. This interaction does not interfere with the $s$-wave driving force of the Weinberg-Tomozawa term, and hence it also goes via the pion-exchange in the box diagram. This box diagram involving the anomalous coupling contributes only to the $VB$ sectors as an attractive potential. When a $VB$ channel develops a bound state, the corresponding $PB$ channel also generates a bound state and the generated state of the $VB$ sector should be heavier than that of the $PB$ sector due to the mass difference of $\bar{D}^*$ and $\bar{D}$. However, thanks to the extra attraction coming from the anomalous term, the $VB$ sector binds more and hence the mixing effect of the $VB$ and $PB$ sectors now becomes more noteworthy.

In this work, we go further with our approach to the $VB$ and $PB$ mixing by implementing the full coupled channels calculation. Indeed, the $VB$ sector can couple to the $PB$ sector via the pion exchange interaction, however this exchange carries momentum transfer and thus in the multiple scattering it is not straightforward to factorize the relevant terms concerning this $VB$-$PB$ mixing. Then we implement the full coupled channels calculation by constructing effective $VB \leftrightarrow PB$ transition potentials from the corresponding box diagrams as done in Ref.~\cite{liang2}

Eventually, six nucleon resonances are generated within the framework of the full coupled channels calculation based on the extended local hidden gauge approach with the anomalous term with several quantum numbers. In addition to the masses and widths of the states, their building blocks or main components are evaluated by the wave function at the origin, as discussed in Ref.~\cite{danijuan}. One of the remarkable observations is the appearance of pairs of orthogonal states mixing the same $PB$ and $VB$ states. In the previous work on open charm baryons \cite{liang2}, in addition to $\Lambda_c (2592)$ which has experimentally been observed, its orthogonal state with an energy $2767$ MeV was also generated. Similarly, in this work two pairs of orthogonal states are also deduced. Moreover, as HQSS is incorporated implicitly within the driving force, the results resemble pairs of heavy quark spin symmetry partners. The meson-baryon hidden charm sector was studied along the same line in Refs.~\cite{wumolina1,wumolina2} but without including pion exchange and hence not mixing the $PB$ and $VB$ states. We shall comment on the analogies and the differences found here with respect to that work. We shall also comment on results of other works that use SU(8) spin-flavour symmetry \cite{carmen} or quark model \cite{lizou}.

\section{Formalism}
\label{sec:fomalism}
\subsection{The interaction via the vector-exchange}
\label{sec:WT}

We study the isospin $I=1/2$ hidden charm states, the $N^*$ resonances with the energy around $4200 \sim 4400$~MeV, following the local hidden gauge approach \cite{hidden1,hidden2,hidden4}. In Refs.~\cite{wumolina2,xiaojuan}, the same states have been investigated within the local hidden gauge approach extended to SU(4). The validity of the use of SU(4) in this kind of studies has been widely discussed in Refs.~\cite{wuzou,gamermannthesis,liang2}. Although SU(4) is broken at the end as a consequence of the multiple scattering and the different masses of the particles within a same multiplet, the elementary vertices which we use here are rather SU(4) symmetric and its use within the present context is acceptable.

We will use the coupled channels formed with anti-charmed mesons $\bar{D}$ and $\bar{D}^*$ and charmed baryons $\Lambda_c,~\Sigma_c,~\Sigma_c^*,~\Lambda_c$, namely $\bar{D}\Lambda_c,~\bar{D}\Sigma_c,~\bar{D}^*\Lambda_c,~\bar{D}^*\Sigma_c,~\bar{D}\Sigma_c^*,~\bar{D}^*\Sigma_c^*$. In Ref.~\cite{wumolina2}, the lighter states such as $\eta_c N,~\pi N,~\eta N,~\eta^{'}N,~ K\Sigma,~K\Lambda$ for the $PB$ sector, were included as a correction (a similar procedure was also considered for the $VB$ sector), however, these states were found not to modify the energies and simply gave around $30$~MeV width to the generated states. In the present work, we do not consider these states but will keep in mind that the states generated that can couple to these states will have a width of that order of magnitude at least.

\begin{figure}[tb]
\epsfig{file=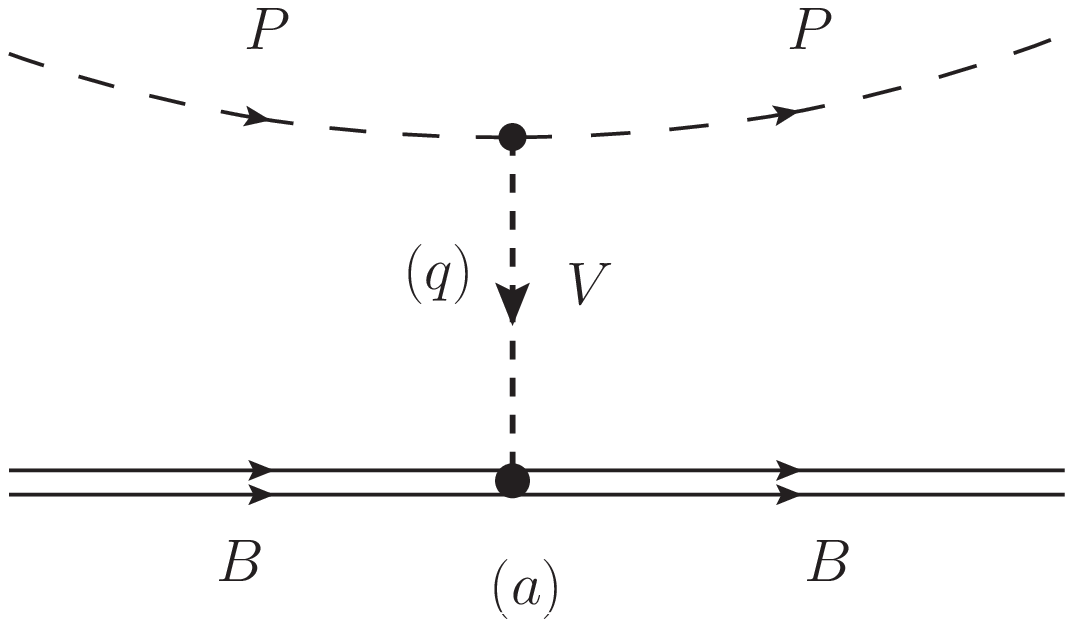, width=5.5cm} ~~~~~\epsfig{file=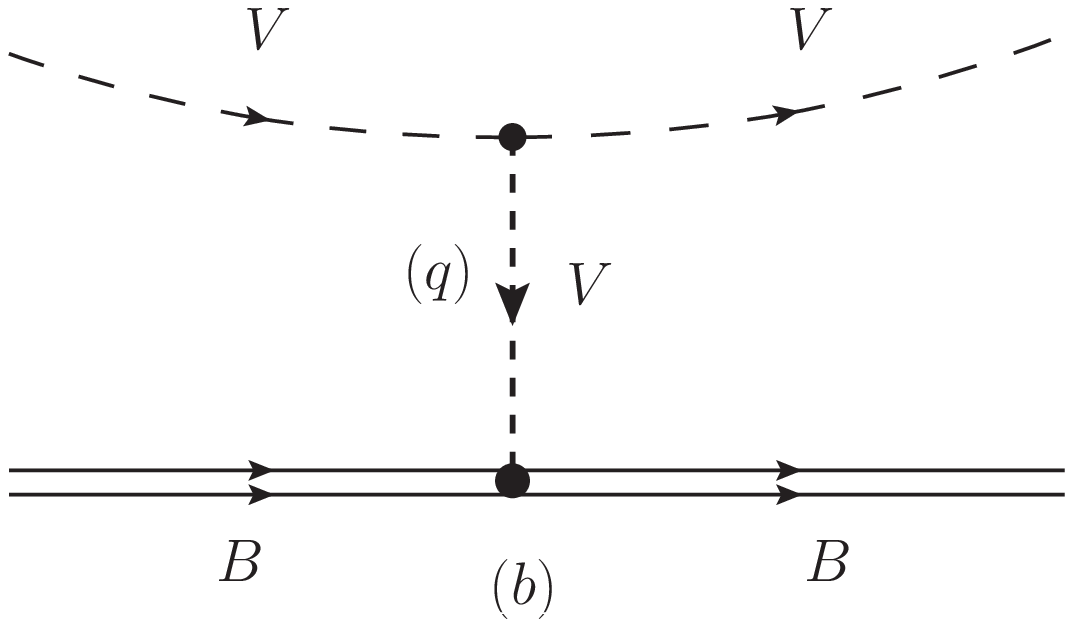, width=5.5cm}
\caption{Diagrammatic representation of the pseudoscalar-baryon interaction (a) and vector-baryon interaction (b).}\label{fig:f1}
\end{figure}

According to the local hidden gauge approach, the meson-baryon interaction proceeds via the exchange of vector mesons as depicted in Fig.~\ref{fig:f1}.
Neglecting the three momenta carried by the exchanged vector, as is automatically done in these studies, we obtain the $s$-wave potential of the $PB$ scattering from channel $i$ to channel $j$ \cite{bennhold}
\begin{equation}
V_{ij} = -C_{ij} \frac{1}{4f^2} (2 \sqrt{s} - M_{B_i} - M_{B_j}) \sqrt{\frac{M_{B_i} + E_i}{2 M_{B_i}}} \sqrt{\frac{M_{B_j} + E_j}{2 M_{B_j}}},
\label{eq:vij}
\end{equation}
with $f=f_\pi=93 \mev$ being the pion decay constant, $M_{B_i}, ~E_i$ ($M_{B_j}, ~E_j$) the mass, energy of baryon of $i$ ($j$) channel. In the case of the $VB$ scattering, the interaction kernels and the following scattering amplitudes have an extra factor $ \vec{\epsilon} \cdot  {\vec{\epsilon}}~'$, the scalar product of the polarization vectors of the incoming and outgoing vector mesons. The coefficients of Eq.~\eqref{eq:vij} are given in Tables~\ref{table:WT_PB8},~\ref{table:WT_VB8},~\ref{table:WT_PB10} and \ref{table:WT_VB10} in the Appendix.

As we shall see in Sec.~\ref{sec:mixing}, in addition to the vector-exchange interaction of Eq. \eqref{eq:vij}, the pion exchange interaction is also taken into account which stems from the local hidden gauge scheme too. This contribution is responsible for mixing the $PB$ and $VB$ sectors and breaking the spin degeneracy as studied in Refs.~\cite{javier,liang,liang2,kanchan1,kanchan2,kanchan3}.

\subsection{The construction of the scattering matrix}
\label{sec:amp}
With the interaction kernel $V_{ij}$ of Eq.~\eqref{eq:vij}, we evaluate the $PB$ or $VB$ two body scattering amplitudes. By solving the Bethe-Salpeter equation in coupled channels, we obtain the scattering amplitude $T$ given in a matrix form
\begin{equation}
T = [1 - V \, G]^{-1}\, V,
\label{eq:Bethe}
\end{equation}
with $G$ being the diagonal loop function which stands for the propagation of the intermediate meson-baryon states. As pointed out in Refs. \cite{xiaooset,wuzou}, the use of the $G$ function with the dimensional regularization scheme can generate unphysical states below the threshold. Hence, in this work, we utilize the $G$ function regularized with the cut off scheme
\begin{equation}
G(s) = \int_0^{q_{\rm max}} \frac{d^{3}\vec{q}}{(2\pi)^{3}}\frac{\omega_P+\omega_B}{2\omega_P\omega_B}\,\frac{2M_{B}}{P^{0\,2}-(\omega_P+\omega_B)^2+i\varepsilon},
\label{eq:Gco}
\end{equation}
where $P^0$ is the CM energy, $s=(P^0)^2$, $\omega_P = \sqrt{\vec{q}\,^2+m_P^2},~\omega_B = \sqrt{\vec{q}\,^2+M_B^2}$, and $q_{\rm max}$ is the cut-off of the three-momentum.

Here, let us refer to a problem which occurs in the cut off method too. The $G$ function with the cut off method also has deficiencies if the cut off is not reasonably larger than the on shell momentum of the intermediate state. This can only affect the $\bar{D}\Lambda_c$ channel in our approach. But the interaction of the $\bar{D}\Lambda_c$ channel is repulsive and does not lead to bound states in this channel. This channel has only a weak and indirect effect in our results which will be discussed below.

\subsection{Breaking the spin degeneracy in vector-baryon sectors}
\label{sec:mixing}

In this subsection we break the spin degeneracy in the $VB$ sector, namely that of the $\bar{D}^*B$ states. Following the approach of Refs.~\cite{javier,revhidden,kanchan1,kanchan2}, we mix states of $\bar{D}^*B_a$ and $\bar{D}B_b$ states in both spin channels as a box diagram such as $\bar{D}^*B_a \to \bar{D}B_b \to \bar{D}^*B_a$ and vice versa, in a first step. Later on we will revert to full coupled channels. The procedure to evaluate these diagrams follows closely what has been done before in the open charm case \cite{liang2}, but in the baryon sector we have more baryons to take into account, while in Ref.~\cite{liang2} only nucleons were involved.

We consider the following $PB \leftrightarrow VB$ mixing processes
   \begin{eqnarray*}
  &&\bar{D}\Sigma_c \to \bar{D}^*\Sigma_c \to \bar{D}\Sigma_c, ~~~~~
  \bar{D}^*\Sigma_c \to \bar{D}\Sigma_c \to \bar{D}^*\Sigma_c, \\
  &&\bar{D}\Sigma_c \to \bar{D}^*\Lambda_c \to \bar{D}\Sigma_c, ~~~~~
  \bar{D}^*\Lambda_c \to \bar{D}\Sigma_c \to \bar{D}^*\Lambda_c, \\
  &&\bar{D}\Sigma_c^* \to \bar{D}^*\Sigma_c^* \to \bar{D}\Sigma_c^*,~~~~~
  \bar{D}^*\Sigma_c^* \to \bar{D}\Sigma_c^* \to \bar{D}^*\Sigma_c^*, \\
  &&\bar{D}^*\Sigma_c \to \bar{D}\Lambda_c \to \bar{D}^*\Sigma_c.
  \end{eqnarray*}
In the mixing we omit the transitions involving the $\pi \Sigma_c \Sigma_c^*$ vertex. The reason is that the SU(4) Clebsch-Gordan coefficient for this vertex is $1/\sqrt{12}$ compared with $1/\sqrt{2}$ for the $\pi N \Delta$. Given the moderate role played by $\pi N \Delta$ transition found in Ref.~\cite{liang}, this approximation is a fair simplifying option. The Clebsch-Gordan coefficient for $\pi \Lambda_c \Sigma_c^*$ is $1/\sqrt{4}$, not so small compared to that of $\pi N \Delta$, but given the fact that the $\bar{D}\Lambda_c$ and $\bar{D}^*\Lambda_c$ channels do not generate bound states by themselves and play only a minor, indirect role in the study, we also neglect terms involving this vertex from the beginning. In addition, note that the $\pi \Lambda_c \Lambda_c$ vertex is zero due to isospin conservation.

\begin{figure}[tb]
\epsfig{file=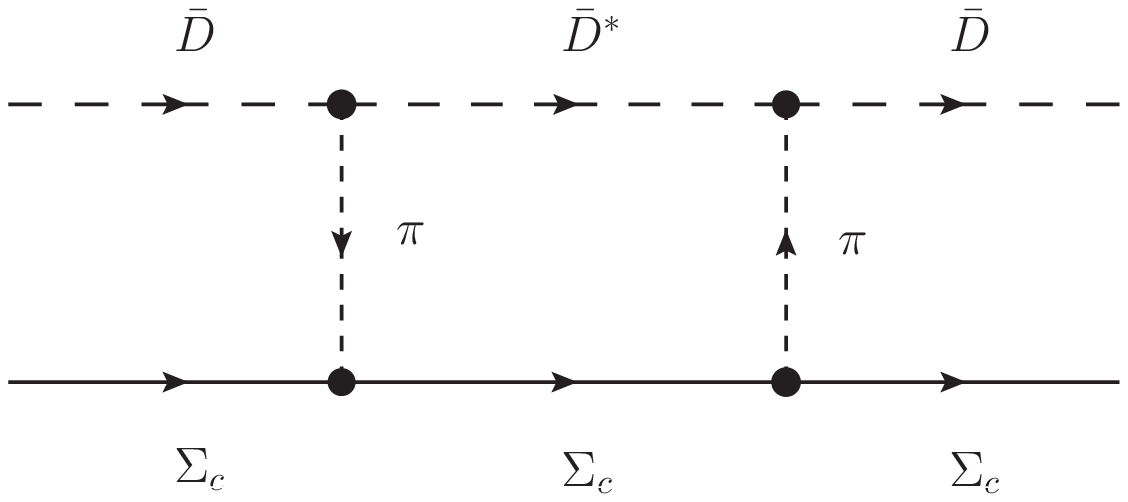, width=6.5cm} ~~~\epsfig{file=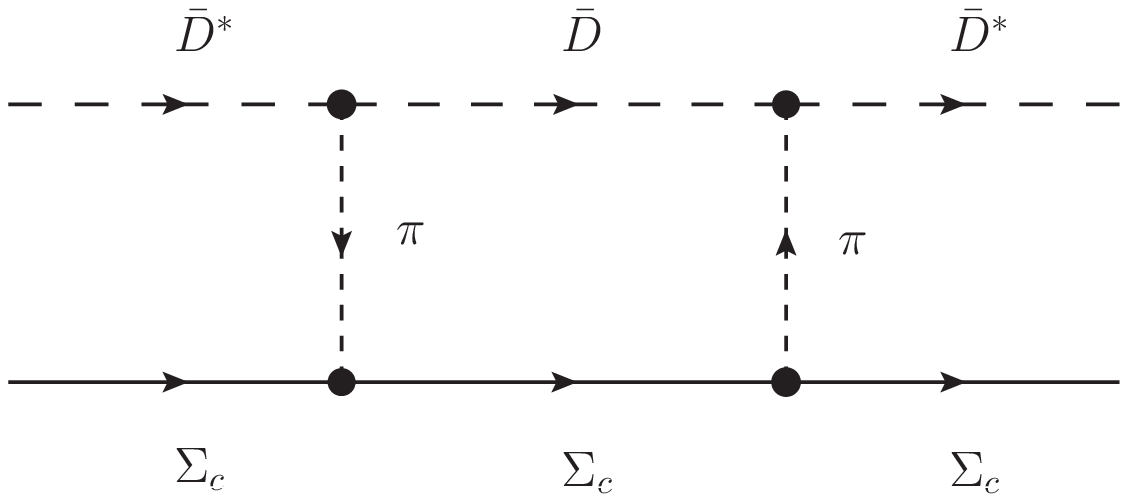, width=6.5cm}
\caption{Diagrammatic representation of the $\bar{D}^* \Sigma_c$ in the intermediate state (left) and the $\bar{D} \Sigma_c$ in the intermediate state (right).}\label{fig:bbbox}
\end{figure}

  In Fig. \ref{fig:bbbox}, a pair of the $\bar{D}\Sigma_c$ and $\bar{D}^*\Sigma_c$ box diagrams, is shown.
As discussed in Sec.~\ref{sec:amp}, we include the $\bar{D}\Lambda_c$ channel only as an intermediate state in the $\bar{D}^*\Sigma_c$ box diagram. Then the channels with $\Sigma_c^*$ are completely separated from the others with $\Sigma_c$ or $\Lambda_c$.

\begin{figure}[tb]
\epsfig{file=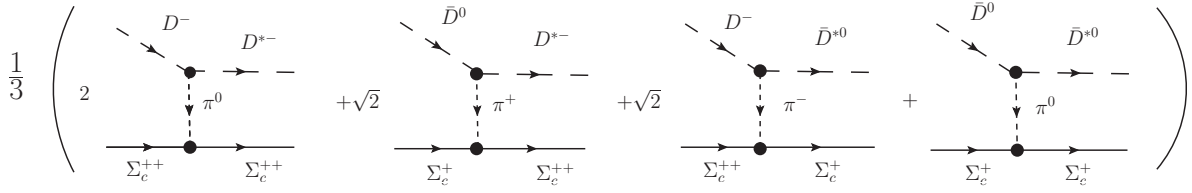, width=16cm}
\caption{Diagrammatic representation of the transition $\bar{D} \Sigma_c \to \bar{D}^* \Sigma_c$ in $I=1/2$.}\label{fig:bbboxhalf}
\end{figure}

First we show the construction of the box potentials with the baryons, $\Lambda_c$ or $\Sigma_c$. We detail the calculation for $\bar{D}\Sigma_c \to \bar{D}^*\Sigma_c \to \bar{D}\Sigma_c$ and use the result as reference to obtain the result in the other cases. By using the $\bar{D} (\bar{D}^*)$ isospin doublet, $(\bar D^0, \, D^-)$ and $\Sigma_c (\Sigma_c^*)$ triplet, $(-\Sigma_c^{++},\Sigma_c^{+},\Sigma_c^{0})$, we write down the relevant hidden charm states with $I=1/2$ in the charge basis
\begin{eqnarray}
|\bar{D} \Sigma_c, \, I=1/2, I_3=+1/2 \rangle
&=&
\sqrt{\frac{2}{3}} |D^- \Sigma_c^{++} \rangle + \sqrt{\frac{1}{3}} |\bar{D}^0 \Sigma_c^+ \rangle, \label{eq:dsig} \\
|\bar{D} \Lambda_c, \, I=1/2, I_3=+1/2 \rangle
&=&
|\bar{D}^0 \Lambda_c^+ \rangle,  \label{eq:dlam}
\end{eqnarray}
and analogously for their excited states such as $\bar{D}^* \Sigma_c,~\bar{D}\Sigma_c^*,~\bar{D}^*\Sigma_c^*$ and $\bar{D}^*\Lambda_c$. In order to evaluate the $\bar{D}\Sigma_c \to \bar{D}^*\Sigma_c$ transition in $I=1/2$, we must consider the diagrams depicted in Fig. \ref{fig:bbboxhalf},
and technically, as also done in Refs.~\cite{liang,liang2}, we evaluate the transitions using only SU(3) symmetry by working in the $(u,d,s)$ sector and using the correspondence of the $(K^+,K^0)$ and $(\bar{K}^0,-K^-)$ isospin doublets with the $(\bar{D}^0,D^-)$ and $(D^+,-D^0)$ ones, respectively. We need the $VPP$ Lagrangian
\begin{equation}
{\cal L}_{VPP} = -ig ~\langle [P,\partial_{\mu}P]V^{\mu}\rangle, \label{eq:vpp}
\end{equation}
where $P,\ V^\mu$ are the ordinary pseudoscalar octet and vector nonet SU(3) matrices of the corresponding fields
\begin{eqnarray}
P &=& \left(
\begin{array}{ccc}
\frac{\pi^0}{\sqrt{2}}+\frac{\eta_8}{\sqrt{6}}  & \pi^+     & K^{+}   \\
\pi^-      & -\frac{\pi^0}{\sqrt{2}}+\frac{\eta_8}{\sqrt{6}} & K^{0}   \\
K^{-}      & \bar{K}^{0}       & -\frac{2\eta_8}{\sqrt{6}}  \\
\end{array}
\right) \
\label{eq:Pmat},  \\
V_\mu &=& \left(
\begin{array}{ccc}
\frac{\rho^0}{\sqrt{2}}+\frac{\omega}{\sqrt{2}} & \rho^+ & \quad K^{*+}  \\
 \rho^{-} & -\frac{\rho^0}{\sqrt{2}} + \frac{\omega}{\sqrt{2}} & K^{*0} \\
  K^{*-} & \bar K^{*0} & \phi \\
\end{array}
\right)_\mu
\label{eq:Vmat}\ .
\end{eqnarray}
and $g=m_V/2f_\pi$ with $m_V \approx 780 \mev$.

From the Lagrangian of Eq.~(\ref{eq:vpp}) with Eqs.~\eqref{eq:Pmat} and \eqref{eq:Vmat}, the $K^* \to K \pi$ decay amplitude that corresponds to the $\bar{D}^* \to \bar{D} \pi$ process is obtained. Yet, in order to get the coupling in the charm sector, it is required to implement an extra factor to account for the different meson field normalizations (1/$\sqrt{2\omega_i}$) and then fulfil the HQSS rules, as pointed out in the previous study of the beauty sector~\cite{liang}. In this case, for instance, the two transition amplitudes, $D^{*-} \to \bar{D}^{0} \pi^-$ and $K^{*0} \to K^{+} \pi^-$, are related as follows
  \begin{equation}
  \frac{t_{\bar{D}^* \to \bar{D}\pi }}{t_{K^* \to K \pi}} \equiv \frac{\sqrt{m_{D^*}m_D}}{\sqrt{m_{K^*}m_K}} \simeq \frac{m_{D^*}}{m_{K^*}}. \label{eq:ratio}
  \end{equation}

\begin{figure}[tb]
\epsfig{file=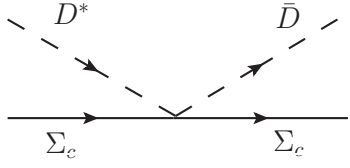, width=5.0cm}
\caption{Diagram of the Kroll Ruderman term.}\label{fig:kroll}
\end{figure}

  The analogy of the strange and charm sectors used before for the $\pi \bar{D}^*\bar{D}$ vertex is also used for the $\pi BB$ vertex, using the correspondence of the isospin multiplets $(-\Sigma^+,~\Sigma^0,~\Sigma_c^-)$ and $(-\Sigma_c^{++},~\Sigma_c^+,~\Sigma_c^0)$. Then we use the standard Yukawa coupling of pions to baryons in SU(3) \cite{pich,ramosphi}, and we obtain the amplitude for the $\bar{D}\Sigma_c \to \bar{D}^*\Sigma_c$ transition 
  \begin{eqnarray}
  -it_{\bar{D}\Sigma_c \to \bar{D}^*\Sigma_c}^P
  &=&
  - \sqrt{2} g \frac{m_{D^*}}{m_{K^*}} (q + P_{\rm in})_\mu \epsilon^\mu \frac{1}{q^2 - m^2_\pi} \frac{2F}{2f_\pi} \vec{\sigma}\,\cdot\, \vec{q} \nonumber \\
  &=&
  2\sqrt{2} g \frac{m_{D^*}}{m_{K^*}} \vec{q}\,\cdot\, \vec{\epsilon} \frac{1}{q^2 - m^2_\pi} \frac{2F}{2f_\pi} \vec{\sigma}\,\cdot\, \vec{q},
  \label{eq:tpi}
\end{eqnarray}
taking $D=0.75$ and $F=0.51$ ~\cite{Borasoy:1998pe} for the two couplings of the Yukawa vertex,
where $P_{\rm in},~P_{\rm out}$ are the incoming, outgoing meson momentum and $q$ is the momentum transfer, and several relations, $P_{\rm in} = q + P_{\rm out}$ and $P_{\rm out} \cdot \epsilon =0$ plus $\epsilon^0 \approx 0$, are used. Note that we work close to the threshold of the vector-baryon channel, which justifies taking $\epsilon^0=0$. In addition to the pion exchange of Fig. \ref{fig:bbboxhalf}, we should also consider the Kroll-Ruderman contact term from the gauge invariance constraint, as depicted in Fig.~\ref{fig:kroll}.
By replacing $\epsilon_\mu (q + P_{\rm in})^\mu \frac{1}{q^2 - m^2_\pi} \vec{\sigma}\,\cdot\, \vec{q}$ of Eq.~\eqref{eq:tpi} by $- \vec{\sigma}\,\cdot\, \vec{\epsilon}$, as discussed in Refs.~\cite{javier,Carrasco:1989vq}, we have the contact term
\begin{eqnarray}
  -it_{\bar{D}\Sigma_c \to \bar{D}^*\Sigma_c}^C
  &=&
  \sqrt{2} g \frac{m_{D^*}}{m_{K^*}} \frac{2F}{2f_\pi} \vec{\sigma}\,\cdot\, \vec{\epsilon}.
  \label{eq:tcon}
  \end{eqnarray}

\begin{figure}[tb]
\epsfig{file=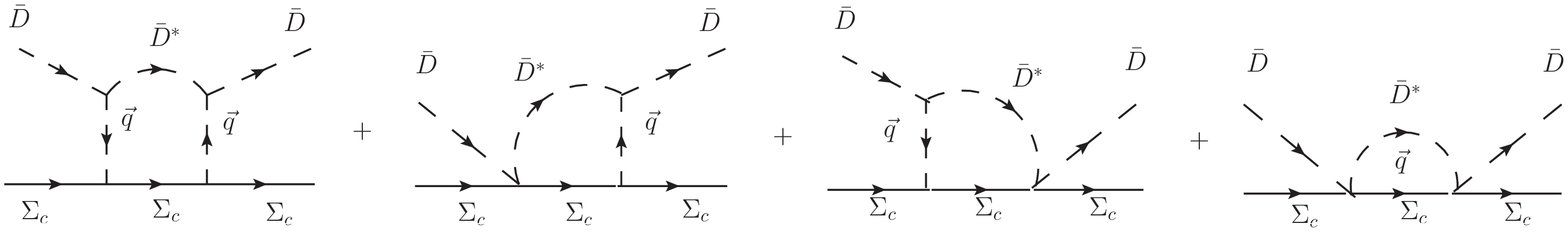, width=16cm}
\caption{The $\bar{D}\Sigma_c \to \bar{D}^*\Sigma_c \to \bar{D}\Sigma_c$ box diagrams.}\label{fig:bbboxtot}
\end{figure}

With the two $\bar{D}\Sigma_c \to \bar{D}^*\Sigma_c$ transition amplitudes of Eqs.~\eqref{eq:tpi} and \eqref{eq:tcon}, we construct the potential for the $\bar{D}\Sigma_c \to \bar{D}^*\Sigma_c \to \bar{D}\Sigma_c $ box diagrams as shown in Fig.~\ref{fig:bbboxtot}, and we obtain
\begin{equation}
\delta V = \delta V^{PP} + 2 \delta V^{PC} + \delta V^{CC}, \label{eq:delv}
\end{equation}
where $\delta V^{PP}$ stands for the first diagram of Fig. \ref{fig:bbboxtot}, $2 \delta V^{PC}$ for the two middle diagrams and $\delta V^{CC}$ for the last one.
As shown in Refs.~\cite{javier,liang,liang2}, we have the box potential of the $\bar{D} \Sigma_c \to \bar{D}^*\Sigma_c \to \bar{D}\Sigma_c$ process
\begin{eqnarray}
-i\delta V^{PP}
&=&
\int \frac{d^4 q}{(2\pi)^4}
\left( \frac{m_{D^*}}{m_{K^*}} \right)^2 \, g \left( 2\sqrt{2} \vec{\epsilon} \,\cdot\,\vec{q} \frac{1}{q^{0\,2} - \vec{q}\,^2 - m^2_\pi} \frac{2F}{2f_\pi} \vec{\sigma}\,\cdot\, \vec{q} \right) \nonumber \\
&& \times (-g) \left( 2\sqrt{2} \vec{\epsilon} \,\cdot\,\vec{q} \frac{1}{q^{0\,2} - \vec{q}\,^2 - m^2_\pi} \frac{2F}{2f_\pi} \vec{\sigma}\,\cdot\, \vec{q} \right) \nonumber \\
&& \times i \frac{1}{2 \omega_{\bar{D}^*}(\vec{q}\,)} \frac{1}{P^0_{\rm in}- q^0 - \omega_{\bar{D}^*}(\vec{q}\,) + i \epsilon} i \frac{M_{\Sigma_c}}{E_{\Sigma_c}(\vec{q}\,)} \frac{1}{K^0_{\rm in}+q^0 - E_{\Sigma_c}(\vec{q}\,) + i \epsilon}, \label{eq:delvpp} \\\nonumber
\end{eqnarray}
\begin{eqnarray}
-i\delta V^{PC}
&=&
\int \frac{d^4 q}{(2\pi)^4} \left( \frac{m_{D^*}}{m_{K^*}} \right)^2 \, g \left( \sqrt{2} \frac{2F}{2f_\pi} \vec{\sigma}\,\cdot\, \vec{\epsilon}  \right)
(-g) \left( 2\sqrt{2} \vec{\epsilon} \,\cdot\,\vec{q} \frac{1}{q^{0\,2} - \vec{q}\,^2 - m^2_\pi} \frac{2F}{2f_\pi} \vec{\sigma}\,\cdot\, \vec{q} \right) \nonumber \\
&& \times i \frac{1}{2 \omega_{\bar{D}^*}(\vec{q}\,)} \frac{1}{P^0_{\rm in}- q^0 - \omega_{\bar{D}^*}(\vec{q}\,) + i \epsilon} i \frac{M_{\Sigma_c}}{E_{\Sigma_c}(\vec{q}\,)} \frac{1}{K^0_{\rm in}+q^0 - E_{\Sigma_c}(\vec{q}\,) + i \epsilon}, \label{eq:delvpc} \\\nonumber
\end{eqnarray}
\begin{eqnarray}
-i\delta V^{CC}
&=&
\int \frac{d^4 q}{(2\pi)^4} \left( \frac{m_{D^*}}{m_{K^*}} \right)^2 \, g \left( \sqrt{2} \frac{2F}{2f_\pi} \vec{\sigma}\,\cdot\, \vec{\epsilon}  \right)
(-g) \left( \sqrt{2} \frac{2F}{2f_\pi} \vec{\sigma}\,\cdot\, \vec{\epsilon}  \right) \nonumber \\
&&
\times i \frac{1}{2 \omega_{\bar{D}^*}(\vec{q}\,)} \frac{1}{P^0_{\rm in}- q^0 - \omega_{\bar{D}^*}(\vec{q}\,) + i \epsilon} i \frac{M_{\Sigma_c}}{E_{\Sigma_c}(\vec{q}\,)} \frac{1}{K^0_{\rm in}+q^0 - E_{\Sigma_c}(\vec{q}\,) + i \epsilon},  \label{eq:delvcc}
\end{eqnarray}
where $P^0_{\rm in}$ and $K^0_{\rm in}$ are the energies of the incoming meson $\bar{D}$ and baryon $\Sigma_c$ respectively, and $\omega_{\bar{D}^*}(\vec{q})=\sqrt{\vec{q}\,^2 +m_{D^*}^2}$, etc., $E_{\Sigma_c}(\vec{q})=\sqrt{\vec{q}\,^2 +m_{\Sigma_c}^2}$.

It is easy to handle the operators $\vec{\sigma}$ and $\vec{\epsilon}$ in the two pion exchange term of Eq.~(\ref{eq:delvpp}) for both cases of the  $PB \to VB \to PB$ and $VB \to PB \to VB$ boxes. In the first case we can use the identity $(\vec{q}\cdot \vec{\sigma})^2=\vec{q}\,^2$, and then the symmetric property in the integral
  \begin{eqnarray}
  q_i q_j \to \frac{1}{3} \vec{q}\,^2 \delta_{ij}
  \label{eq:qsym}
  \end{eqnarray}
and the summation of the polarization of the intermediate vector
  \begin{eqnarray}
  \epsilon_i (\vec{q}) \epsilon_j (\vec{q})
  =
  \delta_{ij}
  \label{eq:pvec} .
  \end{eqnarray}
However, in the $VB \to PB \to VB$ box, the polarization vectors are external and there is no sum over polarizations, one is $\vec{\epsilon}$ and the other $\vec{\epsilon}\,'$. We still can use Eq.~\eqref{eq:qsym} and we obtain in both cases
  \begin{eqnarray}
	  \left\{ \left( \vec{\sigma} \cdot \vec{q} \right) \left( \vec{\epsilon} \cdot \vec{q}  \right) \right\}^2
  &\to&
  \left\{
  \begin{array}{ccc}
  \vec{q}\,^4 & : & PB \to VB \to PB\\
  \frac{1}{3}~\vec{q}\,^4 \vec{\epsilon} \cdot \vec{\epsilon}\,' & : & VB \to PB \to VB \\
  \end{array}
  \right.
  \label{eq:qtotal}.
  \end{eqnarray}
From this result, the inverse box diagram, $\bar{D}^*\Sigma_c \to \bar{D}\Sigma_c \to \bar{D}^*\Sigma_c$, is found to have an extra factor $\frac{1}{3} \vec{\epsilon}\cdot \vec{\epsilon}\,'$. Furthermore, for the contributions which include the Kroll-Ruderman term, namely Eqs.~(\ref{eq:delvpc}) and (\ref{eq:delvcc}), we use again Eq.~\eqref{eq:qsym} plus the property \cite{javier}
  \begin{eqnarray}
  \langle PB | \vec{\sigma} \cdot \vec{\epsilon} | VB \rangle
  =
  \sqrt{3} \delta_{J,1/2}.
  \label{eq:srelation}
  \end{eqnarray}
and we find in both cases that there is only $J=1/2$ contribution and
  \begin{eqnarray}
  (\vec{\sigma} \cdot \vec{\epsilon}) (\vec{\sigma} \cdot \vec{q}) (\vec{\epsilon} \cdot \vec{q})
  =
  \vec{q}\,^2 \delta_{J,1/2}.
  \end{eqnarray}

For other box diagrams involving $\Sigma_c \leftrightarrow \Lambda_c$ baryon conversion, one can obtain the potential in the same manner as done above. The final result is that we can use  Eqs.~\eqref{eq:delvpp}-\eqref{eq:delvcc}, considering the appropriate masses or energies and making the following replacement
  \begin{eqnarray}
  2 \left( \frac{2F}{2f} \right)^2 \to \left( \frac{2D}{2f} \right)^2 .
  \label{eq:pre_REL2}
  \end{eqnarray}

Next, we evaluate the box diagrams of the sectors involving a baryon $\Sigma_c^*$. The two diagrams, $\bar{D}\Sigma_c \to \bar{D}^*\Sigma_c \to \bar{D}\Sigma_c$ and $\bar{D}\Sigma_c^* \to \bar{D}^*\Sigma_c^* \to \bar{D}\Sigma_c^*$, are different in their $\pi BB$ vertices. The $\pi \Sigma_c^* \Sigma_c^*$ vertex is given by
  \begin{eqnarray}
  -it_{\pi \Sigma_c^* \Sigma_c^*}
  =
  \frac{f_{\Sigma^*}}{m_\pi} \vec{S}_{\Sigma^*} \cdot \vec{q}~T_{\Sigma^*},
  \end{eqnarray}
where $\vec{S}_{\Sigma^*}$ and $T_{\Sigma^*}$ represent a spin $S=3/2$ and isospin $I=1$ operator respectively. Using the ordinary quark model, one finds
  \begin{eqnarray}
  \frac{f_{\Sigma_c^*}}{m_\pi}=\frac{4}{5}\frac{f_{\pi NN}}{m_\pi}=\frac{4}{5}\frac{D+F}{2f} .
  \end{eqnarray}
By counting the weight of each contribution in the same manner as for the case of $\Sigma_c$, we have the $\bar{D}\Sigma_c^* \to \bar{D}^* \Sigma_c^*$ transition amplitude
  \begin{eqnarray}
  -it^P_{\bar{D}\Sigma_c^* \to \bar{D}^*\Sigma_c^*}
  =
  2\sqrt{2} g \frac{m_{D^*}}{m_{K^*}} \vec{q}\,\cdot\, \vec{\epsilon} \frac{1}{q^2 - m^2_\pi} \frac{f_{\Sigma^*}}{m_\pi} \vec{S}_{\Sigma_c^*}\,\cdot\, \vec{q}.
  \label{eq:tpi_star}
  \end{eqnarray}
A further simplification is done in this case and we evaluate the box for an average of the diagonal spin transition. We have
  \begin{eqnarray}
  \frac{1}{4} \sum_{m,m^{'}}
  \langle m^{'} | \vec{S}_{\Sigma^*} \cdot \vec{q} | m \rangle
  \langle m | \vec{S}_{\Sigma^*} \cdot \vec{q} | m^{'} \rangle
  =
  \frac{1}{4} q_i q_j \sum_{m,m^{'}}
  \langle m^{'} | S_{\Sigma^*} | m \rangle_i
  \langle m | S_{\Sigma^*} | m^{'} \rangle_j .
  \end{eqnarray}
  Since there is no privileged direction after we sum over $m,m^{'}$ the matrix element, we have
  \begin{eqnarray}
  \frac{1}{4} \sum_{m,m^{'}} \langle m^{'} | S_{\Sigma^*} |m \rangle_i \langle m | S_{\Sigma^*}|m^{'} \rangle_j
  =
  A \delta_{ij} ,
  \end{eqnarray}
and by taking the trace, we obtain the spin factor $A=5/4$. This allows us to write the following relationship
  \begin{eqnarray}
  \left( \vec{S}_{\Sigma^*} \cdot \vec{q} \right) \left( \vec{S}_{\Sigma^*} \cdot \vec{q} \right)
  \to
  \frac{5}{4}\vec{q}\,^2 .
  \end{eqnarray}
Recalling the identity, $(\vec{\sigma}\cdot \vec{q})^2=\vec{q}\,^2$ used in the case of spin $1/2$ baryons in the intermediate states, in order to construct the box potential of the $\bar{D}\Sigma_c^* \to \bar{D}^*\Sigma_c^* \to \bar{D}\Sigma_c^*$, we implement the following substitution in Eqs.~\eqref{eq:delvpp}-\eqref{eq:delvcc}
  \begin{eqnarray}
  \left( \frac{2F}{2f} \right)^2 \to \frac{5}{4} \left( \frac{f_{\Sigma^*}}{m_\pi} \right)^2 .
  \label{eq:sub_decup}
  \end{eqnarray}

\subsection{Contribution from the anomalous term }
\label{sec:anomalous}

\begin{figure}[tb]
\epsfig{file=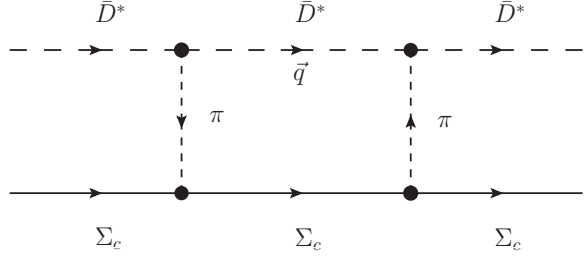, width=8cm}
\caption{Box diagram with anomalous $\bar{D}^* \bar{D}^* \pi$ vertex.}\label{fig:dsnano}
\end{figure}

In addition to the $VB$ and $PB$ mixing term, another correction contributes to the $\bar{D}^*B$ channels which does not interfere with the $s$-wave driving force at tree level. This contribution stems from the anomalous $\bar{D}^*\bar{D}^*\pi$ coupling and is taken into account in a box diagram too. By replacing the $\bar{D}$ mesons in the $PB$-$VB$ box by $\bar{D}^*$, we list down the possible cases for this anomalous box contribution
  \begin{eqnarray*}
  \bar{D}^*\Sigma_c \to \bar{D}^*\Sigma_c \to \bar{D}^*\Sigma_c,& \\
  \bar{D}^*\Sigma_c \to \bar{D}^*\Lambda_c \to \bar{D}^*\Sigma_c,& \\
 \bar{D}^*\Lambda_c \to \bar{D}^*\Sigma_c \to \bar{D}^*\Lambda_c,& \\
  \bar{D}^*\Sigma_c^* \to \bar{D}^*\Sigma_c^* \to \bar{D}^*\Sigma_c^*.
  \end{eqnarray*}
One of them, the $\bar{D}^*\Sigma_c \to \bar{D}^*\Sigma_c \to \bar{D}^*\Sigma_c$ diagram, is depicted in Fig. \ref{fig:dsnano}.

The anomalous $\bar{D}^* \bar{D}^* \pi$ vertex can be obtained from the Lagrangian \footnote{An anomalous process, like the $VVP$ interaction, is one that does not conserve
``natural" parity.
The ``natural" parity of a particle is defined as follows: it is $+1$
if the particle transforms as a true tensor of that rank, and $-1$ if it
transforms as a pseudotensor, e.g. $\pi, \gamma, \rho$ and $a_1$ have ``natural"
parity $-1, +1, +1$ and $-1$, respectively.}
\begin{equation}
{\cal L}_{VVP} = \frac{G'}{\sqrt{2}} \epsilon^{\mu \nu \alpha \beta} \langle \partial_\mu V_\nu \partial_\alpha V_\beta P \rangle,
\end{equation}
with $G' = 3 M_V^2 /16 \pi^2 f^3_\pi (\simeq 14 \gev ^{-1})$ \cite{bramon,luisjose} and the $V$ and $P$ matrices in SU(4) are given in Ref. \cite{danizou}. Thus, for the $\bar{D}^*\Sigma_c^* \to \bar{D}^*\Sigma_c^*$ transition amplitude we obtain
\begin{equation}
-it^{\rm an}_{\bar{D}^*\Sigma_c \to \bar{D}^*\Sigma_c}
=
\frac{1}{2} 2\sqrt{2} \frac{G'}{\sqrt{2}} m_{D^*} \epsilon^{ijk} \epsilon_i (\vec{p})\; q_j \;\epsilon_k (\vec{q}\,) \frac{2F}{2f_\pi} \frac{1}{q^2 - m^2_\pi} \vec{\sigma} \cdot \vec{q}
\label{eq:tpi_ano},
\end{equation}
with a four momentum of the incoming $\bar{D}^*$, $p^\mu \sim (m_{D^*},\vec{0})$, and $\epsilon^{0 \nu \alpha \beta}=\epsilon^{\nu \alpha \beta}$.

In the case of the anomalous term, there is no contact term. Here the box potential is obtained in analogy with the case of the previous box diagram, $\bar{D}\Sigma_c \to \bar{D}^*\Sigma_c \to \bar{D}\Sigma_c$. By using Eqs.~\eqref{eq:qsym} and \eqref{eq:pvec}, the operators of the intermediate states contract and lead to
  \begin{eqnarray}
  \left[ \epsilon^{ijk}\epsilon_i (\vec{p}) q_j \epsilon_k(\vec{q}) \right]^2
  =
  \vec{q}\,^2 \vec{\epsilon} \cdot \vec{\epsilon}~' - (\vec{q} \cdot \vec{\epsilon})(\vec{q} \cdot \vec{\epsilon}~')
  \to
  \frac{2}{3}\vec{q}\,^2 \vec{\epsilon} \cdot \vec{\epsilon}\,' .
  \label{eq:qano}
  \end{eqnarray}
  This, together with $(\vec{\sigma} \cdot \vec{q} )^2=\vec{q}\,^2$ leads to a global factor $\frac{2}{3}\vec{q}\,^4 \vec{\epsilon} \cdot \vec{\epsilon}\,^{'}$ instead of the factor $\frac{1}{3}\vec{q}\,^4 \vec{\epsilon} \cdot \vec{\epsilon}\,^{'}$ that we had in Eq.~\eqref{eq:qtotal} for the case of intermediate vectors.

\section{Full coupled channels calculation}
\label{sec:full}
With an extension of the mixing effect between $PB$ and $VB$ channels caused by the pion exchange, we implement the full coupled channels calculation. Recalling the box potentials constructed in Sec.~\ref{sec:mixing}, we finally study the following four sectors separately.
  \begin{eqnarray*}
  \begin{array}{lll}
    J=1/2 &:& \bar{D}\Sigma_c,~\bar{D}^*\Sigma_c,~\bar{D}^*\Lambda_c \\
    J=3/2&:&\bar{D}^*\Sigma_c,~\bar{D}^*\Lambda_c \\
    J=1/2&:&\bar{D}^*\Sigma_c^*  \\
    J=3/2&:&\bar{D}\Sigma_c^*,~\bar{D}^*\Sigma_c^* \\
  \end{array}
  \end{eqnarray*}
  The driving force, the Weinberg-Tomozawa term, is already given in Sec.~\ref{sec:WT} and it is found that these interactions have no off-diagonal interaction (see Tables~\ref{table:WT_PB8} and \ref{table:WT_VB8}). Furthermore the other interactions, the box diagrams from the local hidden gauge and the anomalous term, are added. Since a large mixture of $VB$ and $PB$ states would be expected from the results of Ref.~\cite{liang2}, it is preferred to implement the full coupled channels calculation, especially between $VB$ and $PB$ channels which generate physical states or have attractive interaction. In this section, first we show the derivation of the effective transition potentials which allow us to implement the full coupled channels calculation, and subsequently the explicit form of the additional interactions to Weinberg-Tomozawa one are given. In Tables~\ref{table:int_B8_1h},~\ref{table:int_B8_3h},~\ref{table:int_B10_1h} and \ref{table:int_B10_3h} of Appendix, the interactions for each sector are summarized.

\subsection{Effective transition potential}
\label{sec:effV}

We follow the scheme of the effective transition potential constructed in Ref.~\cite{liang2}. In this work, the generated states are looked for in the $s$-wave scattering amplitude, where the Weinberg-Tomozawa interaction appears. However, the intermediate states of the box potentials are not necessarily in the $s$-wave and then the box potentials should be decomposed into two parts to implement full coupled channels scheme. The part with intermediate states in the $s$-wave is utilized to construct the $s$-wave effective transition potential and the rest part is added to the Weinberg-Tomozawa interactions as a correction. On the other hand, since the anomalous $\bar{D}^*\bar{D}^*\pi$ couping does not interfere with the Weinberg-Tomozawa term, these terms contribute only via the box diagram.

With the $s$-wave component of the $PB_a \to VB_b \to PB_a$ box potential, $\delta V_{PB_a \to VB_b \to PB_a}$($s$-wave), such as $\bar{D}\Sigma_c \to \bar{D}^*\Lambda_c (s\textrm{-wave}) \to \bar{D}\Sigma_c$, we define an effective transition potential between $PB_a$ and $VB_b$ channels as
%Fig. \ref{fig:bbboxtot}
\begin{equation}
	\tilde{V}^2_{\rm eff} G_{VB_b} = \delta V_{PB_a \to VB_b \to PB_a} (s\textrm{-wave}), \label{eq:eff1}
\end{equation}
with $G_{VB_b}$ being the $G$ function of the intermediate $VB_b$ channel.
In the same manner, from the counterpart diagram, $VB_b \to PB_a \to VB_b$, we can also define the other $PB_a \leftrightarrow VB_b$ transition potential with a corresponding box potential $\delta{V}^{'}$
\begin{equation}
	\tilde{V}^{'\,2}_{\rm eff} G_{PB_a} = \delta V{'}_{VB_b \to PB_a \to VB_b} (s\textrm{-wave}). \label{eq:eff2}
\end{equation}
To ensure the symmetric property of the potential or the amplitude matrices, namely $V_{ij}=V_{ji}$, we define the effective transition potential by taking the average of the two potentials
\begin{equation}
V_{\rm eff} = \frac{1}{2} (\tilde{V}_{\rm eff} + \tilde{V}'_{\rm eff}). \label{eq:eff}
\end{equation}
However, it turns out that the potentials are very similar and in fact they are identical in the limit of $m_{D}=m_{D^*}$ as demonstrated in Refs.~\cite{liang,liang2}, hence, the concept of $V_{\rm eff}$ is well defined. From the definitions, $\tilde{V}_{\rm eff}$ and $\tilde{V}_{\rm eff}^{'}$ are found to be double-valued functions. Therefore, the relative signs of the two functions should be taken equal not to be canceled out. We choose $\tilde{V}_{\rm eff}$ with negative real part as it corresponds to a virtual pion exchange. The energy of the generated states does not depend upon this prescription.

Among the relevant box diagrams, it is obvious that the term involving the contact term, Eqs.~\eqref{eq:delvpc} and \eqref{eq:delvcc}, contribute only to the $s$-wave due the product $\vec{\sigma} \cdot \vec{\epsilon}$. Therefore, the intermediate states of these terms are always in the $s$-wave. On the other hand, the case of the two pion exchange term is different. When the extraction of the $s$-wave component is necessary, we utilize the following separation for each of the pion exchange
  \begin{eqnarray}
  \left( \vec{\sigma}\cdot \vec{q} \right) \left( \vec{\epsilon} \cdot \vec{q} \right)
  \to
  \epsilon_i q_i \sigma_j q_j
  =
  \epsilon_i \sigma_j
  \left\{
  \frac{1}{3}q^2 \delta_{ij} + \left( q_i q_j - \frac{1}{3}q^2 \delta_{ij} \right)
  \right\}
  \label{eq:pre_qs},
  \end{eqnarray}
where the first term of the last expression of the equation corresponds to the $s$-wave component and the second to $d$-wave, and this separation leads to
  \begin{eqnarray}
  \left\{ \left( \vec{\sigma} \cdot \vec{q} \right) \left( \vec{\epsilon} \cdot \vec{q}  \right) \right\}^2 (s\textrm{-wave})
  &\to&
  \left\{
  \begin{array}{ccc}
  \frac{1}{3} \vec{q}\,^4 & : & PB \to VB \to PB\\
  \frac{1}{3} \frac{1}{3}~\vec{q}\,^4 \vec{\epsilon} \cdot \vec{\epsilon}\,' & : & VB \to PB \to VB \\
  \end{array}
  \right.
  \label{eq:qs}.
  \end{eqnarray}
Subtracting the $s$-wave contribution from the total contribution of the box we obtain the $d$-wave contribution.

In the following sections, the explicit forms of the potentials are shown where the expressions of $I^{'}_1, I^{'}_2, I^{'}_3, FAC, I_1, I_2,$ and $I_3$ can be seen in Eqs. (34)-(37) and (41)-(43) of Ref.~\cite{liang}.

\subsubsection{$\bar{D}\Sigma_c, \bar{D}^*\Sigma_c, \bar{D}^*\Lambda_c$ with $J=1/2$}
\label{sec:B8_1h}

In this sector, the $VB$ intermediate states in the $PB \to VB \to PB$ box potentials can have $s$-wave and $d$-wave contributions. By using the decomposition, Eq.~\eqref{eq:qs}, and recalling that the contact-contact and contact-pion terms contribute to only the $s$-wave component, we can decompose them into two parts
  \begin{eqnarray}
  \delta V_I (s\textrm{-wave})
  &=&
  \delta V
  \left( \bar{D}\Sigma_c \to \bar{D}^*\Sigma_c \to \bar{D}\Sigma_c; J=1/2, s\textrm{-wave} \right) \nonumber \\
  &=&
  REL1 \times FAC \times \left(\frac{1}{3}\frac{\partial I_1}{\partial m_{\pi}^2} + 2 I_2 + I_3 \right)
  \label{eq:pvp1s}, \\
  \delta V_{II} (s\textrm{-wave})
  &=&
  \delta V
  \left( \bar{D}\Sigma_c \to \bar{D}^*\Lambda_c \to \bar{D}\Sigma_c; J=1/2, s\textrm{-wave} \right) \nonumber \\
  &=&
  REL2 \times FAC \times \left(\frac{1}{3}\frac{\partial I_1}{\partial m_{\pi}^2} + 2 I_2 + I_3 \right)
  \label{eq:pvp2s},
  \end{eqnarray}
  \begin{eqnarray}
  \delta V
  \left( \bar{D}\Sigma_c \to \bar{D}^*\Sigma_c \to \bar{D}\Sigma_c; J=1/2, d\textrm{-wave} \right)
  &=&
  REL1 \times FAC \times \frac{2}{3}\frac{\partial I_1}{\partial m_{\pi}^2}
  \label{eq:pvp1d}, \\
  \delta V
  \left( \bar{D}\Sigma_c \to \bar{D}^*\Lambda_c \to \bar{D}\Sigma_c; J=1/2, d\textrm{-wave} \right)
  &=&
  REL2 \times FAC \times \frac{2}{3}\frac{\partial I_1}{\partial m_{\pi}^2}
  \label{eq:pvp2d},
  \end{eqnarray}
where we use the factor $FAC$ together with the relative factors as discussed in Sec.~\ref{sec:mixing}
  \begin{eqnarray}
  FAC
  &=&
  \frac{9}{2} g^2 \left( \frac{m_{D^*}}{m_{K^*}} \right)^2 \left( \frac{F+D}{2f} \right)^2 , \\
  REL1
  &=&
  \frac{4}{9}
  \left( \frac{2F}{D+F} \right)^2,
  \label{eq:rel1} \\
  REL2
  &=&
  \frac{1}{9}
  \left( \frac{2D}{D+F} \right)^2 .
  \label{eq:rel2}
  \end{eqnarray}

The $PB$ intermediate states in the $VB \to PB \to VB$ box potentials are necessarily in the $s$-wave and thus with Eqs.~\eqref{eq:qtotal}, \eqref{eq:rel1} and \eqref{eq:rel2} we have
  \begin{eqnarray}
  \delta V^{'}_{I} (s\textrm{-wave})
  &=&
  \delta V
  \left( \bar{D}^* \Sigma_c \to \bar{D} \Sigma_c \to \bar{D}^* \Sigma_c; J=1/2 \right) \nonumber \\
  &=&
  REL1 \times FAC \times \left(\frac{\partial I_1^{'}}{\partial m_{\pi}^2} + 2 I_2^{'} + I_3^{'} \right)
  \label{eq:vpv1_1h}, \\
  \delta V^{'}_{II} (s\textrm{-wave})
  &=&
  \delta V
  \left( \bar{D}^* \Lambda_c \to \bar{D} \Sigma_c \to \bar{D}^* \Lambda_c; J=1/2 \right) \nonumber \\
  &=&
  REL2 \times FAC \times \left(\frac{\partial I_1^{'}}{\partial m_{\pi}^2} + 2 I_2^{'} + I_3^{'} \right)
  \label{eq:vpv3_1h}.
  \end{eqnarray}

From the $s$-wave box potentials given by Eqs.~\eqref{eq:pvp1s} and \eqref{eq:vpv1_1h} and by Eqs.~\eqref{eq:pvp2s} and \eqref{eq:vpv3_1h}, we construct the following two effective transition potentials
  \begin{eqnarray}
  V^{\rm eff}_{\bar{D}\Sigma_c \leftrightarrow \bar{D}^*\Sigma_c}
  &=&
  \frac{ \sqrt{\delta V_{I} (s\textrm{-wave})/G_{\bar{D}^*\Sigma_c} } + \sqrt{ \delta V_{I}^{'} (s\textrm{-wave})/G_{\bar{D}\Sigma_c} }}{2}
  \label{eq:veff1}, \\
  V^{\rm eff}_{\bar{D}\Sigma_c \leftrightarrow \bar{D}^*\Lambda_c}
  &=&
  \frac{\sqrt{\delta V_{II} (s\textrm{-wave})/G_{\bar{D}^*\Lambda_c}} + \sqrt{\delta V_{II}^{'} (s\textrm{-wave})/G_{\bar{D}\Sigma_c}}}{2}
  \label{eq:veff2}.
  \end{eqnarray}
They are placed in the corresponding off-diagonal channels.

Furthermore, as discussed before, the $\bar{D}\Lambda_c$ contribution is just added to the $\bar{D}^*\Sigma_c$ channel as a box correction,
\begin{eqnarray}
  \delta V
  \left( \bar{D}^* \Sigma_c \to \bar{D} \Lambda_c \to \bar{D}^* \Sigma_c; J=1/2 \right)
  &=&
  REL2 \times FAC \times \left(\frac{\partial I_1^{'}}{\partial m_{\pi}^2} + 2 I_2^{'} + I_3^{'} \right)
  \label{eq:vpv3}.
\end{eqnarray}

We also show the box potentials which stem from the anomalous term
  \begin{eqnarray}
  \delta V_{\rm an}
  \left( \bar{D}^* \Sigma_c \to \bar{D}^* \Sigma_c \to \bar{D}^* \Sigma_c \right)
  &=&
  REL1 \times AFAC \times \frac{\partial I_1^{'}}{\partial m_{\pi}^2}
  \label{eq:ano1}, \\
  \delta V_{\rm an}
  \left( \bar{D}^* \Sigma_c \to \bar{D}^* \Lambda_c \to \bar{D}^* \Sigma_c \right)
  &=&
  REL2 \times AFAC \times \frac{\partial I_1^{'}}{\partial m_{\pi}^2}
  \label{eq:ano2}, \\
  \delta V_{\rm an}
  \left( \bar{D}^* \Lambda_c \to \bar{D}^* \Sigma_c \to \bar{D}^* \Lambda_c \right)
  &=&
  REL2 \times AFAC \times \frac{\partial I_1^{'}}{\partial m_{\pi}^2}
  \label{eq:ano3},
  \end{eqnarray}
with a factor which comes from Eqs.~\eqref{eq:tpi_ano} and \eqref{eq:qano}
  \begin{eqnarray}
  AFAC
  &=&
  \frac{9}{8} G^{'2} \left( \frac{D+F}{2f} \right)^2 m_{D^*}^2 .
  \end{eqnarray}
These box potentials are added to the $\bar{D}^*\Sigma_c$ and $\bar{D}^*\Lambda_c$ diagonal Weinberg-Tomozawa interaction equally in the $J=1/2$ and $J=3/2$ sectors.

\subsubsection{$\bar{D}^*\Sigma_c, \bar{D}^*\Lambda_c$ with $J=3/2$}
\label{sec:B8_3h}

Since the $PB$ intermediate states with $J=3/2$ are automatically in the $d$-wave, we just add the following box diagram potentials to the diagonal interactions
  \begin{eqnarray}
  \delta V
  \left( \bar{D}^* \Sigma_c \to \bar{D}\Sigma_c \to \bar{D}^* \Sigma_c; J=3/2 \right)
  &=&
  REL1 \times FAC \times \frac{\partial I_1^{'}}{\partial m_{\pi}^2}
  \label{eq:vpv1_3h}, \\
  \delta V
  \left( \bar{D}^* \Lambda_c \to \bar{D} \Sigma_c \to \bar{D}^* \Lambda_c; J=3/2 \right)
  &=&
  REL2 \times FAC \times \frac{\partial I_1^{'}}{\partial m_{\pi}^2}.
  \label{eq:vpv2_3h}, \\
  \delta V
  \left( \bar{D}^* \Sigma_c \to \bar{D}\Lambda_c \to \bar{D}^* \Sigma_c; J=3/2 \right)
  &=&
  REL2 \times FAC \times \frac{\partial I_1^{'}}{\partial m_{\pi}^2}
  \label{eq:vpv3_3h}.
  \end{eqnarray}
Remember that the box potentials from the anomalous term, Eqs.~\eqref{eq:ano1}-\eqref{eq:ano3}, are also added to this sector. Note that in this sector there is no transition between the $\bar{D}^*\Sigma_c$ and $\bar{D}^*\Lambda_c$ channels and hence, in this sector we do not have coupled channels but two single channels.

\subsubsection{$\bar{D}^*\Sigma_c^*$ with $J=1/2,~5/2$}
\label{sec:B10_1h}

In this sector, we have only a $\bar{D}^*\Sigma_c^*$ channel. Considering that the $\bar{D}\Sigma_c^*$ channel with $J=1/2,~5/2$ can not be in the $s$-wave, we find that the box potentials should include only the pion-pion term with Eq.~\eqref{eq:qtotal}. Taking into account Eq.~\eqref{eq:sub_decup}, we have the $\bar{D}^*\Sigma_c^* \to \bar{D}\Sigma_c^* \to \bar{D}^*\Sigma_c^*$ box diagram
  \begin{eqnarray}
  \delta V
  \left( \bar{D}^* \Sigma_c^* \to \bar{D} \Sigma_c^* \to \bar{D}^* \Sigma_c^*; J=1/2 \right)
  &=&
  REL3 \times FAC \times \frac{\partial I_1^{'}}{\partial m_{\pi}^2}
  \label{eq:vpv4_1h},
  \end{eqnarray}
with
  \begin{eqnarray}
  REL3
  &=&
  \frac{5}{9}
  \left( \frac{f_{\Sigma^*}}{m_{\pi}} \right)^2/ \left( \frac{D+F}{2f} \right)^2
  =
  \frac{16}{45}.
  \end{eqnarray}
This is just added to the $\bar{D}^*\Sigma_c^*$ diagonal Weinberg-Tomozawa interaction with the extra contribution from the anomalous term
  \begin{eqnarray}
  \delta V_{\rm an}
  \left( \bar{D}^* \Sigma_c^* \to \bar{D}^* \Sigma_c^* \to \bar{D}^* \Sigma_c^* \right)
  &=&
  REL3 \times AFAC \times \frac{\partial I_1^{'}}{\partial m_{\pi}^2}
  \label{eq:ano4}.
  \end{eqnarray}

\subsubsection{$\bar{D}\Sigma_c^*$ and $\bar{D}^*\Sigma_c^*$ with $J=3/2$}
\label{sec:B10_3h}

We here consider the $\bar{D}\Sigma_c^*$ and $\bar{D}^*\Sigma_c^*$ coupled channels. The $\bar{D}\Sigma_c^*$ intermediate state with $J=3/2$ has an admixture of the $s$-wave and $d$-wave components and thus each potential is separated into $s$- and $d$-wave contributions as done in the $PB$-$VB$ sector with $J=1/2$
  \begin{eqnarray}
	  \delta V_{III}^{'} (s\textrm{-wave})
  &=&
  \delta V
  \left( \bar{D}^* \Sigma_c^* \to \bar{D} \Sigma_c^* \to \bar{D}^* \Sigma_c^*; J=3/2, s\textrm{-wave} \right) \nonumber \\
  &=&
  REL3 \times FAC \times \left( \frac{1}{3}\frac{\partial I_1^{'}}{\partial m_{\pi}^2} + 2 I_2^{'} + I_3^{'} \right)
  \label{eq:vpv4s_3h},
  \end{eqnarray}
and
  \begin{eqnarray}
  \delta V
  \left( \bar{D}^* \Sigma_c^* \to \bar{D} \Sigma_c^* \to \bar{D}^* \Sigma_c^*; J=3/2, d\textrm{-wave} \right)
  &=&
  REL3 \times FAC \times \frac{2}{3}\frac{\partial I_1^{'}}{\partial m_{\pi}^2}
  \label{eq:vpv4d_3h}.
  \end{eqnarray}
On the other hand, although the $\bar{D}^*\Sigma_c^*$ intermediate state of the $\bar{D}\Sigma_c^* \to \bar{D}^*\Sigma_c^* \to \bar{D}\Sigma_c^*$ box diagram also has the $s$-wave component, the many spin contributions of $\bar{D}^*\Sigma_c^*$ complicate the analysis. Therefore, we adopt only the definition of Eq.~\eqref{eq:eff2} instead of Eq.~\eqref{eq:eff} to construct the $\bar{D}\Sigma_c^* \leftrightarrow \bar{D}^*\Sigma_c^*$ effective transition potential
  \begin{eqnarray}
  V^{\rm eff}_{\bar{D}\Sigma_c^* \leftrightarrow \bar{D}^*\Sigma_c^*}
  &=&
  \sqrt{\delta V_{III}^{'} (s\textrm{-wave})/G_{\bar{D}\Sigma_c^*}}
  \label{eq:veff3},
  \end{eqnarray}
and by using this effective potential, we extract the $s$-wave contribution from the total $\bar{D}\Sigma_c^* \to \bar{D}^*\Sigma_c^* \to \bar{D}\Sigma_c^*$ box
  \begin{eqnarray}
  \delta V
  \left( \bar{D} \Sigma_c^* \to \bar{D}^* \Sigma_c^* \to \bar{D} \Sigma_c^*; J=3/2, {\rm except}~ s\textrm{-wave} \right) \nonumber \\
  =
  REL3 \times FAC \times \left( \frac{\partial I_1 }{\partial m_{\pi}^2} + 2I_2 + I_3 \right)
  - \left( V^{\rm eff}_{\bar{D}\Sigma_c^* \leftrightarrow \bar{D}^*\Sigma_c^*} \right)^2 G_{\bar{D}^*\Sigma_c^*}
  \label{eq:pvp4}.
  \end{eqnarray}
We also add the potential from the anomalous term, Eq.~\eqref{eq:ano4}, to the $\bar{D}^*\Sigma_c^*$ channel.

\subsection{Resonance measured on the real axis}
\label{sec:resonance}

In the full coupled channels scheme, we would see the generation of several physical states. Under the coupled channels unitary scheme, the corresponding poles are looked for in the second Riemann sheet in the complex plane. In the case of bound states, poles appear on the real axis below the threshold in the first Riemann sheet and thus we look for the bound states poles as usual. From the residue of the scattering amplitude at the pole, $M_R$, we obtain the coupling constant $g_i,~g_j$
  \begin{eqnarray}
  g_ig_j = \lim_{\sqrt{s}\to M_R} (\sqrt{s}-M_R)T_{ij}(\sqrt{s}).
  \end{eqnarray}
With the coupling constant and the $G$ function at the bound state pole, we evaluate the wave function at the origin of channel $i$ as $g_iG_i(M_R)$ \cite{danijuan}.

Resonance poles, on the other hand, emerge in the complex energy plane. As we do not know how to extend the box potentials or effective potentials to the complex energy plane, we analyze the resonance states in another way. Instead of looking for a pole position of the scattering amplitude, we search a peak position $M_R$ on the real axis, where the square of the magnitude of the amplitude $|T(\sqrt{s})|^2$ has a maximum. Since this position is not exactly the same between different scattering channels, we determine it in the dominant channel. Together with the peak position, we measure the width of the state, $\Gamma_R$, on the real axis as being a distance between two points where $|T(\sqrt{s})|^2$ has a half maximum value.

Next, we determine the coupling constants of resonance states with the measured mass and width. On the real axis and in the vicinity of the peak position, we assume the Breit-Wigner amplitude
  \begin{eqnarray}
  T_{ij}(\sqrt{s})
  \sim
  \frac{g_i g_j}{\sqrt{s}-M_R+i\Gamma_R/2}.
  \end{eqnarray}
At the peak position, the coupling constants of the dominant channel $d$ is given by the imaginary part of the amplitude
  \begin{eqnarray}
  {\rm Im} T_{ij} (M_R)
  =
  - \frac{g_i g_j}{\Gamma_R/2} ,~
  g_d
  =
  \sqrt{\left| \frac{\Gamma_R}{2}{\rm Im}T_{dd}(M_R) \right|},
  \label{eq:coupling1}
  \end{eqnarray}
where $g_d$ is taken as a positive number. In order not to lose the relative signs, the coupling constants of the other channels are determined from the ratio of the imaginary part of the amplitude in different channels
  \begin{eqnarray}
  g_i
  =
  \frac{{\rm Im}T_{id}(M_R)}{{\rm Im}T_{dd}(M_R)} g_d .
  \label{eq:coupling2}
  \end{eqnarray}
The wave function at the origin, or the probability, can be obtained from the coupling constant and the $G$ function at the pole position. In this case, since we determine the coupling at the peak position, for the sake of consistency we evaluate the wave function at the origin of channel $i$ as $g_i G_i(M_R)$.

\section{Numerical results of the full coupled channels calculation}
In this section, we show the results of the full coupled channels approach. In order to obtain the numerical results, we take three cut off parameters, $q_{\rm max}^P$ in the pseudoscalar-baryon loop, $q_{\rm max}^V$ in the vector-baryon loop and $q_{\rm max}^B$ in the box diagrams. Since no nucleon resonance state in the high energy region of our interest has been reported, we tentatively adopt the cut off parameters used in the previous work of open charm baryons \cite{liang2} where the parameters were chosen to reproduce two $\Lambda_c$ resonances, $\Lambda_c (2592)$ and $\Lambda_c(2625)$. In Table~\ref{table:cutoff}, three sets of three cut off parameters are listed. Furthermore, as well as in the previous work, we put the Yukawa form factor, $F(\vec{q}\,) = \Lambda^2 / (\Lambda^2 + \vec{q}\,^2)$, on the $\pi B B'$ vertices in all the box diagrams, with cut off $\Lambda = 1000 \mev$, a standard value. In the following subsections, we show the results of the $\bar{D}\Sigma_c$ and $\bar{D}^* \Sigma_c$ states and $\bar{D}\Sigma_c^*$ and $\bar{D}^*\Sigma_c^*$ states separately.

%%% Cut off parameters %%%
\begin{table}[H]
     \renewcommand{\arraystretch}{1.5}
     \setlength{\tabcolsep}{0.2cm}
\centering
   \begin{tabular}{|c|c|c|c|}
	   \hline
              & set I & set II &  set III \\
	   \hline
   $q_{\rm max}^B$ & 600 & 800 &  1000 \\
   $q_{\rm max}^V$ & 771 & 737 &  715  \\
   $q_{\rm max}^P$ & 527 & 500 &  483 \\
   \hline
   \end{tabular}
   \caption{Cut off parameters, $q_{\rm max}^{V} (q_{\rm max}^P)$ for the vector-baryon (pseudoscalar-baryon) $G$ function, $q_{\rm max}^B$ for all the box potentials.}
\label{table:cutoff}
\end{table}

%%% Sigma_c states %%%
\subsection{$\bar{D}\Sigma_c$ and $\bar{D}^*\Sigma_c$ states}

In Table~\ref{table:peak_1hBO_22}, the peak positions and the widths of the resonant states with $J=1/2$, generated by the coupled channels of $\bar{D}\Sigma_c$, $\bar{D}^*\Sigma_c$ and $\bar{D}^*\Lambda_c$ with $\bar{D}\Lambda_c$ box diagrams, are listed. As discussed in Sec.~\ref{sec:resonance}, in this work, we extract the nature of the resonant states not by looking for the corresponding poles in the complex energy plane but by measuring the scattering amplitude on the real axis. We determine the peak position and width of generated states from the $\bar{D}^*\Sigma_c \to \bar{D}^*\Sigma_c$ scattering amplitude.

We obtain two different narrow resonances. With the ambiguities from the choice of the cut off parameters, the lower energy state appears around 4228 MeV $\pm$ 15 MeV with about 20 MeV width, while the higher energy state is very close to the $\bar{D}\Lambda_c$ threshold, 4295 MeV with 10 MeV width. The coupling constants $g_i$ and the wave functions at the origin $g_iG_i$ for the middle set of the cut off parameters, obtained from Eqs.~\eqref{eq:coupling1} and \eqref{eq:coupling2}, are listed in Table~\ref{tab:g_1hBO_2}. From these wave functions, it is found that two states appear as two orthogonal states roughly as $\frac{1}{\sqrt{2}} \left( \bar{D}\Sigma_c \pm \bar{D}^*\Sigma_c \right)$. This situation is similar to the one observed in Ref.~\cite{liang2} where the $\frac{1}{\sqrt{2}}(DN - D^*N)$ corresponded to the $\Lambda_c(2595)$, while there was a prediction for an orthogonal state about $\frac{1}{\sqrt{2}}(DN+D^*N)$ which was much less bound \footnote{In Ref.~\cite{liang2} the sign of $V_{\rm eff}$ was taken positive and here we take it negative. The energies do not depend upon this prescription but the $\frac{1}{\sqrt{2}}(DN \pm D^*N)$ combination would correspond now to $\frac{1}{\sqrt{2}} (DN \mp D^*N)$.}. This feature was also observed in Ref.~\cite{Romanets:2012hm} where the higher energy state appears with zero binding \cite{juantalk}.

It is also remarkable that for both orthogonal states, the $\bar{D}^*\Sigma_c$ component has a bit stronger coupling than the $\bar{D}\Sigma_c$ component. This could be one of the consequence of the heavy quark limit. Although, even in the charm sector, the mass difference between $D$ and $D^*$, $m_{D^*}-m_{D} \sim 140$~MeV is not so small yet, the vector-baryon system binds more than the pseudoscalar-baryon due to the anomalous term contributions and this mechanism makes it possible that the vector-baryon and pseudoscalar-baryon systems couple easier.

%%% Peaks of Dstar-bar Sigma_c with J=1/2 %%%
\begin{table}[h]
     \renewcommand{\arraystretch}{1.5}
     \setlength{\tabcolsep}{0.2cm}
\centering
   \begin{tabular}{|c|c|c|c|}
   \hline
     & set~I & set~II &  set~III \\
   \hline
   peak 1   & $4241.7$ & $4227.6$ & $4218.6$   \\
   width 1  & $19.5$   & $21.1$ & $21.5$   \\
   \hline
   peak 2   & $4296.8$ & $4295.1$ & $4294.5$   \\
   width 2  & $13.1$ & $10.6$ & $9.6$   \\
   \hline
   \end{tabular}
   \caption{Peak positions and their widths in the full coupled channels of $\bar{D} \Sigma_c[4321],~\bar{D}^*\Sigma_c[4462],~\bar{D}^*\Lambda_c[4295]$ with $J=1/2$ measured in the $\bar{D}^*\Sigma_c \to \bar{D}^*\Sigma_c$ channel as a function of $q_{\rm max}^{B,V,P}$. (The number in brackets after the channel indicates the threshold mass of the channel. Units: MeV)}
\label{table:peak_1hBO_22}
\end{table}

%%% coupling set II %%%
\begin{table}[ht]
\renewcommand{\arraystretch}{1.2}
\centering
\begin{tabular}{cccc}
\hline\hline
   $(4227.6, 21.1)$ & $ \bar{D} \Sigma_c $ & $\bar{D}^* \Sigma_c$ & $\bar{D}^* \Lambda_c$\\
   \hline
   $g_i$  & $4.40$ & $5.39$ & $0.39$ \\
   $g_i\,G_i$ & $-15.66$ & $-24.17$ & $-3.31$ \\
   \hline
   $(4295.1, 10.6)$ & $ \bar{D} \Sigma_c $ & $\bar{D}^* \Sigma_c$ & $\bar{D}^* \Lambda_c$ \\
   \hline
   $g_i$  & $-1.27$ & $2.28$ & $-0.11$ \\
   $g_i\,G_i$ & $8.46$ & $-12.60$ & $2.09+i0.04$ \\
   \hline
\end{tabular}
\caption{The coupling constants to various channels for the poles in the $J=1/2$ sector of $\bar{D}^* \Sigma_c$ and coupled channels, taking the cutoff Set II of Table~\ref{table:cutoff}.}
\label{tab:g_1hBO_2}
\end{table}

Next we see the $\bar{D}^*\Sigma_c$ and $\bar{D}^*\Lambda_c$ single channels with $J=3/2$. As expected, the $\bar{D}^*\Lambda_c$ channel generates no state and we observe the $\bar{D}^*\Sigma_c$ broad resonance whose mass is 4217 MeV and width is around 100 MeV as listed in Table~\ref{table:peak_3hBO}. This  resonance is generated below the $\bar{D}\Sigma_c^*$ threshold but can decay into $\bar{D}\Lambda_c$ via the pion exchange, as well as the two resonance states with $J=1/2$. Compared to $J=1/2$ resonances, this $J=3/2$ resonance is much broader because of the difference of the $\bar{D}^*\Sigma_c \to \bar{D}\Lambda_c \to \bar{D}^*\Sigma_c$ box diagram as seen in Eqs.~\eqref{eq:vpv3_1h} and \eqref{eq:vpv3_3h}.

Apart from masses or widths of the generated states, it is remarkable the similarity between the results of the hidden and open charm baryon. The vector-baryon system whose interaction is attractive, such as in $\bar{D}^*\Sigma_c$ or $D^*N$, generate deeply bound states with $J=1/2$ and $J=3/2$ and the pion exchange removes their spin degeneracies. On the other hand, the corresponding pseudoscalar-baryon system such as $\bar{D}\Sigma_c$ or $DN$ generates shallow-bound states with $J=1/2$. As a consequence, in the $J=1/2$ sector, a pair of orthogonal states emerges as a mixture of $VB$ and $PB$ bound states, while in the $J=3/2$ sector one deeply (quasi-)bound state is generated.

%%% peaks of Dstar-bar Sigma_c with J=3/2 %%%
\begin{table}[h]
     \renewcommand{\arraystretch}{1.5}
     \setlength{\tabcolsep}{0.2cm}
\centering
   \begin{tabular}{|c|c|c|c|}
   \hline
     & set~I & set~II &  set~III \\
   \hline
   peak   & $4250.5$ & $4217.7$ & $4205.8$   \\
   width  & $140.8$ & $103.2$ & $82.0$   \\
   \hline
   \end{tabular}
   \caption{Peak positions and their widths in the $\bar{D}^* \Sigma_c [4462]$ single channel with $J=3/2$ sector as a function of $q_{\rm max}^{B,V,P}$. (Units: MeV)}
\label{table:peak_3hBO}
\end{table}
%\caption{The coupling constants to various channels for the poles in the $I=0, \ J^P=3/2^-$ sector of $D^* N$ and coupled channels, with the anomalous term and taking $q_{max}^{B,V}=800,737$ MeV.  In bold face we highlight the main components.}

We come back here to the results of Refs.~\cite{wumolina1,wumolina2} and admit an extra width of about $30$ MeV for the two lower $\frac{1}{\sqrt{2}}(\bar{D}^*\Sigma_c \pm \bar{D}\Sigma_c)$ states from the coupling to the light $PB$ or $VB$ sectors. 

\subsection{$\bar{D}\Sigma_c^*$ and $\bar{D}^*\Sigma_c^*$ states}

In this subsection we show the results of the generated states involving $\bar{D}\Sigma_c^*$ and $\bar{D}^* \Sigma_c^*$ channels. In these sectors, bound states are developed and then we look for the corresponding poles as discussed in Sec.~\ref{sec:resonance}. In Table~\ref{table:pole_1hBD}, the pole positions of states generated by the $\bar{D}^*\Sigma_c^*$ single channel with $J=1/2,5/2$ are listed. We obtain a $\bar{D}^*\Sigma_c^*$ bound state with an energy $4344$ with $10$ MeV uncertainty from the cutoff setup, which appears around $180$ MeV below the $\bar{D}^*\Sigma_c^*$ threshold.

%%% peaks of Dstar-bar Sigma_c-star with J=1/2 %%%
\begin{table}[h]
     \renewcommand{\arraystretch}{1.5}
     \setlength{\tabcolsep}{0.2cm}
\centering
   \begin{tabular}{|c|c|c|c|}
   \hline
          & set~I & set~II &  set~III \\
   \hline
   Pole     & $4354.5$ & $4344.1$ & $4337.5$   \\
  \hline
   \end{tabular}
\caption{Poles in the $\bar{D}^* \Sigma_c^*[4527]$ single channel with $J=1/2,~5/2$ as a function of cut off $q_{\rm max}^{B,V,P}$. (Units: MeV)}
\label{table:pole_1hBD}
\end{table}

Next we see the results of coupled channels of $\bar{D}\Sigma_c^*$ and $\bar{D}^* \Sigma_c^*$ with $J=3/2$. The pole positions of the generated states are listed in Table~\ref{table:pole_3hBD}. In this sector, we obtain two bound states which are not far from the $\bar{D}^*\Sigma_c^*$ bound states with $J=1/2$. The two states are separated by about $50$ MeV and the higher state appears a few MeV below the $\bar{D}\Sigma_c^*$ threshold. We show the coupling constants and the wave functions at the origin with the cut off set II in Table~\ref{tab:g_3hBD_2}. From these wave functions, it is found that two bound states are the mixture of $\bar{D}\Sigma_c^*$ and $\bar{D}^*\Sigma_c^*$ corresponding roughly to the  $\frac{1}{\sqrt{2}}(\bar{D}\Sigma_c^* \pm \bar{D}^*\Sigma_c^*)$ combinations.

%%% peaks of Dstar-bar Sigma_c-star with J=3/2 %%%
  \begin{table}[h]
  \renewcommand{\arraystretch}{1.5}
  \setlength{\tabcolsep}{0.2cm}
  \centering
  \begin{tabular}{|c|c|c|c|}
  \hline
             & set~I & set~II &  set~III \\
  \hline
  Pole 1 & $4330.6$ & $4324.9$ & $4319.9$   \\
  Pole 2 & $4384.1$ & $4377.8$ & $4374.4$   \\
  \hline
  \end{tabular}
  \caption{Poles in coupled channels of $\bar{D} \Sigma_c^*[4385],~ \bar{D}^*\Sigma_c^* [4527]$ with $J=3/2$ as a function of $q_{\rm max}^{B,V,P}$. (Units: MeV)}
  \label{table:pole_3hBD}
  \end{table}

%%% coupling set II %%%
\begin{table}[ht]
\renewcommand{\arraystretch}{1.2}
\centering
\begin{tabular}{ccc}
\hline\hline
$4324.85$ & $ \bar{D} \Sigma_c^* $ & $\bar{D}^* \Sigma_c^*$ \\
   \hline
   $g_i$  & $3.61$ & $4.89$ \\
   $g_i\,G_i^{II}$ & $-16.54$ & $-24.22$ \\
   \hline
$4378.84$ & $ \bar{D} \Sigma_c^* $ & $\bar{D}^* \Sigma_c^*$ \\
   \hline
   $g_i$  & $-1.25$ & $3.00$ \\
   $g_i\,G_i^{II}$ & $12.30$ & $-17.57$ \\
   \hline
\end{tabular}
\caption{The coupling constants to various channels for the poles in the $J=3/2$ sector of $\bar{D} \Sigma_c^*,~\bar{D}^*\Sigma_c^*$, taking the cutoff Set II of Table~\ref{table:cutoff}.}
\label{tab:g_3hBD_2}
\end{table}

\section{Summary and conclusions}

In this work, the dynamics of an anti-charmed meson and a charmed baryon is studied in order to investigate the hidden charm resonances with isospin $I=1/2$, $N^*$, in the energy region around $4200 \sim 4400$ MeV. Since all the hadronic particles of our interest contain the charm quark, the extended local hidden gauge approach with SU(4) symmetry, as studied in Refs.~\cite{wumolina2,xiaojuan,xiaooset}, is employed to extract the information of the relevant vertices. Consequently, the leading interaction proceeds via the vector exchange and results in the Weinberg-Tomozawa interaction. Furthermore, we also take into account two types of additional interactions, as a box diagram correction. One of them with the $\bar{D}\bar{D}^*\pi$ vertex from the local hidden gauge scheme, proceeds via the pion exchange and requires the corresponding contact term. The other one comes from the anomalous $\bar{D}^*\bar{D}^* \pi$ interaction. The former additional interaction allows the $PB$ and $VB$ sectors to couple to each other. As pointed out in the previous works, we also place high importance on this mixing effect. Hence we have solved the Bethe-Salpeter equation under the full coupled channels scheme by constructing the effective transition potential following the discussion of Ref.~\cite{liang2}. The particular values of the $\pi BB$ vertices, allowed us to classify the states into four sectors of the hidden charm systems, $\bar{D}\Sigma_c, \bar{D}^*\Sigma_c^*, \bar{D}^*\Lambda_c$ with $J=1/2$, $\bar{D}^*\Sigma_c, \bar{D}^*\Lambda_c^*$ with $J=3/2$, $\bar{D}^*\Sigma_c^*$ with $J=1/2$ and $\bar{D}\Sigma_c^*, \bar{D}^*\Sigma_c^*$ with $J=3/2$.

By looking for the bound state poles and measuring the resonant peaks on the real axis, as discussed in Sec.~\ref{sec:resonance}, we have seen that six states appear with several angular momenta. Furthermore, from the wave function at the origin, the dominant components of the states have been evaluated. The properties of the states, masses, widths, dominant components and main decay channel, are summarized in Table \ref{tab:polesum}.
\begin{table}[H]
     \renewcommand{\arraystretch}{1.5}
     \setlength{\tabcolsep}{0.3cm}
\centering
\begin{tabular}{|c|c|c|c|}
\hline
main channel & $J$   &   $\, (E, \, \Gamma)\ \text{[MeV]} \,$  &  main decay channels  \\
\hline
$\frac{1}{\sqrt{2}}(\bar{D}^* \Sigma_c + \bar{D} \Sigma_c)$ & $1/2$   & $4228, \, 21(51)$  & $\bar{D} \Lambda_c$ \\
\hline
$\frac{1}{\sqrt{2}}(\bar{D}^*\Sigma_c - \bar{D} \Sigma_c)$  & $1/2$ & $4295, \, 11(41)$ & $\bar{D} \Lambda_c$ \\
\hline
$\bar{D}^* \Sigma_c$  &  $3/2$  & $4218, \, 103$ & $\bar{D} \Lambda_c$\\
\hline
$\bar{D}^* \Sigma_c^*$  &  $1/2,~5/2$ & $4344, \, 0$ & $-$\\
\hline
$\frac{1}{\sqrt{2}}(\bar{D}^* \Sigma_c^* + \bar{D} \Sigma_c^*)$ & $3/2$   & $4325, \, 0$  & $ - $ \\
\hline
$\frac{1}{\sqrt{2}}(\bar{D}^*\Sigma_c^* - \bar{D} \Sigma_c^*)$  & $3/2$ & $4378, \, 0$ &$-$ \\
\hline
\end{tabular}
\caption{Energies and widths of the obtained states with the dominant component and main decay channels of each state. All the states are nucleon resonances with negative parity and have an estimated uncertainty of $\pm 20 \mev$. The numbers in brackets for the first two states correspond to the estimated width, adding the $30$ MeV width obtained in Refs.~\cite{wumolina1,wumolina2} from coupling to the light $PB$ or $VB$ sectors.}
\label{tab:polesum}
\end{table}

The three generated states with $\Sigma_c$ are, a pair of orthogonal states with $J=1/2$, $\frac{1}{\sqrt{2}}(\bar{D}^* \Sigma_c \pm \bar{D} \Sigma_c)$, and, a resonance dominated by the $\bar{D}^*\Sigma_c$ component with $J=3/2$. The two orthogonal states with energies around $4228$ and $4295$ MeV respectively, differ from each other by about $65$ MeV and have a small width due to the weak coupling to the $\bar{D}\Lambda_c$. On the other hand, the $\bar{D}^*\Sigma_c$ state with $J=3/2$ emerges around $4218$ MeV with a large width $103$ MeV. It is instructive to see the similarity in the generated states with $\Sigma_c^*$. Around $100$ MeV above the region of the three state of $\Sigma_c$, a pair of the orthogonal states with $J=3/2$, $\frac{1}{\sqrt{2}} (\bar{D}^*\Sigma_c^* \pm \bar{D}\Sigma_c^*)$ and one spin degenerate state with $J=1/2,5/2$ of $\bar{D}\Sigma_c^*$ develop. Considering the mass difference between $\Sigma_c$ and $\Sigma_c^*$, $M_{\Sigma_c^*}-M_{\Sigma_c}\sim 65$ MeV, this result might indicate the realization of the heavy quark spin symmetry regarding the charmed baryon pair, $\Sigma_c$ and $\Sigma_c^*$.

In the end, let us discuss the result comparing them to those of Refs.~\cite{wumolina2,xiaojuan,xiaooset}. Apart from the use of the different cutoff parameters and the neglect of the light hadron sectors that simply give the width to the generated states, we employ the same interactions for the $PB$ and $VB$ sectors respectively. In Ref.~\cite{wumolina2}, the $PB$ sectors with the same quantum number develop physical states around $4250$ MeV, not so far away from our results, however, the $VB$ states appear much higher than ours, about $200$ MeV higher. This is also the case in the work done in Ref.~\cite{zouli} done along the same lines.

In all these works \cite{xiaojuan,xiaooset,wumolina2}, the box diagrams were not considered and the $PB$ and $VB$ sectors were studied independently. We have seen in the present work that due to the pion exchange that induces $PB$ and $VB$ mixing, the $VB$ states get much bound. In addition, the pion exchange in the anomalous sector, which is only operative for the $VB$ sector, provides a further contribution to the $VB$ binding.

Another work that deserves attention in this discussion is the one of Ref.~\cite{carmen}. There a unitarized coupled channel approach, also mixing $PB$ and $VB$ states is done. The authors use the Weinberg-Tomozawa interaction as leading contribution and an extended SU(8) spin-flavour symmetry constraint, with modifications implemented to respect HQSS. They obtain also bound states but the energies are almost $200$ MeV smaller than what we find in the present work.

There are other calculations based on the quark model \cite{lizou}. The authors also find bound states for different spins, and the energies range between $3900 \sim 4500$ MeV. One should note that in this latter work, the predictions differ by about, $200 \sim 300$ MeV depending on the interaction used, a color magnetic interaction in one case and a chiral interaction in the other. A detailed discussion of these different model is done in Ref.~\cite{juantalk}.

The comparison of the results obtained here with those of other works is instructive and indicate the large uncertainties in the predictions in this sector. In this respect, one should mention an earlier work where states are found with a binding of $1000$ MeV \cite{hofmann}, which, as discussed in Ref.~\cite{wumolina2}, is somewhat extreme. These large theoretical differences contrast with the situation in other sectors, like the one of open charm, where the constraints of reproducing existing experimental data makes the predictions of different approaches more similar. Yet, one thing is shared by all these approaches and this is the existence of many bound states of the $N^*$ type.

The former discussion should be sufficient incentive to search for these states, which would bring a clear evidence of exotic baryonic states, with a quark structure significantly different from the standard one of the baryons made up of three quarks.

\section*{Acknowledgments}

This work is partly supported by the Spanish Ministerio de Economia y Competitividad and European FEDER funds under Contract No. FIS2011-28853-C02-01 and the Generalitat Valenciana in the program Prometeo, 2009/090. We acknowledge the support of the European Community-Research Infrastructure Integrating Activity Study of Strongly Interacting Matter (Hadron Physics 3, Grant No. 283286) under the Seventh Framework Programme of the European Union.
This work is also partly supported by the National Natural Science Foundation of China under Grant No. 11165005.

\section*{Appendix: Summary of the interaction}
\begin{appendix}

In this appendix we write the $C_{ij}$ coefficients of Eq.~\eqref{eq:vij} for the different channels.

\setcounter{table}{0}
\renewcommand{\thetable}{A\arabic{table}}

\begin{table}[H]
     \renewcommand{\arraystretch}{1.5}
     \setlength{\tabcolsep}{0.4cm}
\centering
\begin{tabular}{c|c}
	$C_{ij}$ & $\bar{D}\Sigma_c$  \\
\hline
$\bar{D} \Sigma_c$  &  1 \\
\end{tabular}
\caption{$C_{ij}$ coefficients for $\bar{D} \Sigma_c$ with $I=1/2$ and $J^P=1/2^-$.}
\label{table:WT_PB8}
\end{table}

\begin{table}[H]
     \renewcommand{\arraystretch}{1.5}
     \setlength{\tabcolsep}{0.4cm}
\centering
\begin{tabular}{c|cc}
	$C_{ij}$ & $\bar{D}^*\Sigma_c$ &  $\bar{D}^* \Lambda_c$  \\
\hline
$\bar{D}^* \Sigma_c$  &  1 & 0  \\
$\bar{D}^* \Lambda_c$ &    & -1  \\
\end{tabular}
\caption{$C_{ij}$ coefficients for $\bar{D}^* \Sigma_c$ and $\bar{D}^*\Lambda_c$ with $I=1/2$ and $J^P=1/2^-,~3/2^-$.}
\label{table:WT_VB8}
\end{table}

\begin{table}[H]
     \renewcommand{\arraystretch}{1.5}
     \setlength{\tabcolsep}{0.4cm}
\centering
\begin{tabular}{c|c}
	$C_{ij}$ & $\bar{D}\Sigma_c^*$  \\
\hline
$\bar{D} \Sigma_c^*$  &  1 \\
\end{tabular}
\caption{$C_{ij}$ coefficients for $\bar{D} \Sigma_c^*$ with $I=1/2$ and $J^P=3/2^-$.}
\label{table:WT_PB10}
\end{table}

\begin{table}[H]
     \renewcommand{\arraystretch}{1.5}
     \setlength{\tabcolsep}{0.4cm}
\centering
\begin{tabular}{c|c}
	$C_{ij}$ & $\bar{D}^*\Sigma_c^*$  \\
\hline
$\bar{D}^* \Sigma_c^*$  &  1 \\
\end{tabular}
\caption{$C_{ij}$ coefficients for $\bar{D}^* \Sigma_c^*$ with $I=1/2$ and $J^P=1/2^-,~3/2^-,~5/2^-$.}
\label{table:WT_VB10}
\end{table}

\begin{table}[H]
     \renewcommand{\arraystretch}{1.5}
     \setlength{\tabcolsep}{0.4cm}
\centering
\begin{tabular}{c|ccc}
	& $\bar{D}\Sigma_c$ & $\bar{D}^*\Sigma_c$ & $\bar{D}^*\Lambda_c$  \\
\hline
$\bar{D} \Sigma_c$    &  Eqs.~\{ \eqref{eq:vij} + \eqref{eq:pvp1d} + \eqref{eq:pvp2d} \} & Eq.~\eqref{eq:veff1} & Eq.~\eqref{eq:veff2} \\
$\bar{D}^* \Sigma_c$  &   & Eqs.~\{ \eqref{eq:vij} + \eqref{eq:vpv3} + \eqref{eq:ano1} + \eqref{eq:ano2} \} & 0 \\
$\bar{D}^* \Lambda_c$ &   &   & Eqs.~\{\eqref{eq:vij} + \eqref{eq:ano3} \} \\
\end{tabular}
\caption{Interactions in the coupled channels $\bar{D}\Sigma_c, \bar{D}^*\Sigma_c, \bar{D}^*\Lambda_c$ with $J^P=1/2^-$.}
\label{table:int_B8_1h}
\end{table}

\begin{table}[H]
     \renewcommand{\arraystretch}{1.5}
     \setlength{\tabcolsep}{0.4cm}
\centering
\begin{tabular}{c|cc}
	& $\bar{D}^*\Sigma_c$ & $\bar{D}^*\Lambda_c$  \\
\hline
$\bar{D}^* \Sigma_c$    &  Eqs.~\{ \eqref{eq:vij} + \eqref{eq:vpv1_3h} + \eqref{eq:vpv3_3h} + \eqref{eq:ano1} + \eqref{eq:ano2} \} & 0 \\
 $\bar{D}^* \Lambda_c$  &   & Eqs.~\{ \eqref{eq:vij} + \eqref{eq:vpv2_3h} + \eqref{eq:ano3} \} \\
\end{tabular}
\caption{Interactions in the two single channels $\bar{D}^*\Sigma_c, \bar{D}^*\Lambda_c$ with $J^P=3/2^-$.}
\label{table:int_B8_3h}
\end{table}

\begin{table}[H]
     \renewcommand{\arraystretch}{1.5}
     \setlength{\tabcolsep}{0.4cm}
\centering
\begin{tabular}{c|c}
	& $\bar{D}^*\Sigma_c^*$  \\
\hline
$\bar{D}^* \Sigma_c^*$    &  Eqs.~\{ \eqref{eq:vij} + \eqref{eq:vpv4_1h} + \eqref{eq:ano4} \} \\
\end{tabular}
\caption{Interactions in the single channel $\bar{D}^*\Sigma_c^*$ with $J^P=1/2^-,~5/2^-$.}
\label{table:int_B10_1h}
\end{table}

\begin{table}[H]
     \renewcommand{\arraystretch}{1.5}
     \setlength{\tabcolsep}{0.4cm}
\centering
\begin{tabular}{c|cc}
	& $\bar{D}\Sigma_c^*$ & $\bar{D}^*\Sigma_c^*$  \\
\hline
$\bar{D} \Sigma_c^*$    &  Eqs.~\{ \eqref{eq:vij} + \eqref{eq:pvp4} \} &  Eq.~\eqref{eq:veff3}  \\
$\bar{D}^* \Sigma_c^*$  &   & Eqs.~\{ \eqref{eq:vij} + \eqref{eq:vpv4d_3h} + \eqref{eq:ano4} \} \\
\end{tabular}
\caption{Interactions in the coupled channels $\bar{D}\Sigma_c^*, \bar{D}^*\Sigma_c^*$ with $J^P=3/2^-$.}
\label{table:int_B10_3h}
\end{table}

\end{appendix}

\bibliographystyle{ursrt}

%\end{spacing}
\end{document}